\documentclass[manuscript,screen]{acmart}
\usepackage{tikz}
\usepackage{caption}
\usepackage{subcaption}
\usepackage{subcaption}
\usepackage{natbib}
\usepackage{bchart}
\usepackage{graphicx}
\usepackage{longtable}
\usepackage{multirow}
\usepackage{float}
\usetikzlibrary{mindmap}\usepackage{listings}
\AtBeginDocument{%
  \providecommand\BibTeX{{%
    \normalfont B\kern-0.5em{\scshape i\kern-0.25em b}\kern-0.8em\TeX}}}

\setcopyright{acmcopyright}
\copyrightyear{2023}
\acmYear{2023}
\acmDOI{XXXXXXX.XXXXXXX}

\acmConference[ACM '23]{Make sure to enter the correct
  conference title from your rights confirmation emai}{
  2023}{ACM}
\acmPrice{15.00}
\acmISBN{978-1-4503-XXXX-X/18/06}

\begin{document}

\title[A Systematic Review on Fostering Appropriate Trust in Human-AI Interaction]{A Systematic Review on Fostering Appropriate Trust in Human-AI Interaction: Trends, Opportunities and Challenges}

\author{Siddharth Mehrotra}
\email{s.mehrotra@tudelft.nl}
\orcid{0000-0002-2067-3451}
\affiliation{%
  \institution{Delft University of Technology}
  \streetaddress{Van Mourik Broekmanweg 6}
  \city{Delft}
  \state{Zuid Holland}
  \country{The Netherlands}
  \postcode{2628XE}
}

\author{Chadha Degachi}
\affiliation{%
 \institution{Delft University of Technology}
  \streetaddress{Van Mourik Broekmanweg 6}
  \city{Delft}
  \state{Zuid Holland}
  \country{The Netherlands}
  \postcode{2628XE}}

\author{Oleksandra Vereschak}
\affiliation{%
 \institution{Sorbonne Université, ISIR}
 \country{France}}

\author{Catholijn M. Jonker}
\affiliation{%
 \institution{Delft University of Technology}
  \streetaddress{Van Mourik Broekmanweg 6}
  \city{Delft}
  \state{Zuid Holland}
  \country{The Netherlands}
  \postcode{2628XE}}

\author{Myrthe L. Tielman}
\affiliation{%
 \institution{Delft University of Technology}
  \streetaddress{Van Mourik Broekmanweg 6}
  \city{Delft}
  \state{Zuid Holland}
  \country{The Netherlands}
  \postcode{2628XE}}

\renewcommand{\shortauthors}{Mehrotra et al.}

\begin{abstract}
  Appropriate Trust in Artificial Intelligence (AI) systems has rapidly become an important area of focus for both researchers and practitioners. Various approaches have been used to achieve it, such as confidence scores, explanations, trustworthiness cues, or uncertainty communication. However, a comprehensive understanding of the field is lacking due to the diversity of perspectives arising from various backgrounds that influence it and the lack of a single definition for appropriate trust. To investigate this topic, this paper presents a systematic review to identify current practices in building appropriate trust, different ways to measure it, types of tasks used, and potential challenges associated with it. We also propose a Belief, Intentions, and Actions (BIA) mapping to study commonalities and differences in the concepts related to appropriate trust by (a) describing the existing disagreements on defining appropriate trust, and (b) providing an overview of the concepts and definitions related to appropriate trust in AI from the existing literature. Finally, the challenges identified in studying appropriate trust are discussed, and observations are summarized as current trends, potential gaps, and research opportunities for future work.  Overall, the paper provides insights into the complex concept of appropriate trust in human-AI interaction and presents research opportunities to advance our understanding on this topic.
\end{abstract}



\maketitle

\section{Introduction}
Artificial Intelligence (AI) has become an increasingly ubiquitous technology in recent years, with applications in a wide range of industries and areas of life. The ability of AI to process and analyze large amounts of data quickly and accurately makes it particularly valuable for domains with high-stake decision-making such as finance, healthcare, and transportation \cite{sharma2020artificial,mou2019artificial}. 
While AI-embedded systems are powerful, they can still fail or behave unpredictably, leading to inappropriate trust, and introducing the corresponding risk of \textit{misuse} and \textit{disuse}  \cite{parasuraman1997humans}. 

Both disuse \cite{marsh2005trust} and misuse \cite{mcbride2010trust} of AI-embedded systems by humans have led to severe issues, such as Amazon's AI recruiting tool being biased against women \cite{dastin2018amazon}, a railroad accident in which the crew neglected speed constraints \cite{sorkin1988likelihood}, and the use of facial recognition technology in law enforcement to target Black and Latino communities \cite{jones2020law}. One of the major reasons of disuse and misuse of AI is people's over- or under-trust in it, or in other words, lack of appropriate trust in AI \cite{ososky2013building}. Appropriate trust is often linked to the alignment between the perceived and actual performance of the system \cite{yang2020visual}. We argue that human trust in the AI system must be appropriate because, with appropriate trust in AI, people may be simultaneously aware of the potential and the limitations of AI. This should lead to reducing the harms and negative consequences of misuse and disuse of AI \cite{parasuraman2003automation}. 

People have long been aware of the importance of appropriate trust in interpersonal relationships \cite{eisenstadt1984patrons}. Taking an example from the Indian scripture ``\textit{Bhagavad Gita}'', dated 400 BCE \cite{flood1996introduction}, the deity \textit{Krishna} advises that humans should be careful in trusting others and develop trust in degrees so that their trust is often appropriate \cite{burke_2016}. Furthermore, he suggests by cultivating appropriate trust, humans gradually move forward in interpersonal relationships. This highlights for how long this concept has played a role and is vital and helpful in understanding how people can develop appropriate trust in interpersonal relationships and AI systems. 

To achieve appropriate trust in AI systems, different approaches have been taken such as use of confidence scores \cite{zhang2020confidence,kaniarasu2013robot,bansal2021does,poursabzi2021manipulating, ma2023correctnesstrust}, explanations \cite{wang2021explanations,lai2020chicago, lai2019human, sivaraman2023ignoretrust}, cues (alarms \cite{chen2021automation,yang2017uxtransparnecy}, warning signals \cite{okamura2020adaptive} or uncertainty communication \cite{tomsett2020rapid}). Many studies aim to adjust the trust bestowed in a system to reflect the trustworthiness of said system \cite{zhang2020confidence,liu2021interactiveexplain,schlicker2021towards, yang2023biolitcalibration}. Despite these efforts, a comprehensive understanding of the field is currently lacking, and consensus on the definition of appropriate trust remains elusive. Different perspectives and varying definitions of trust, trustworthiness, and reliance contribute to this lack of clarity, as pointed out by Gille \cite{gille2020we}. 

According to Jacovi et al.'s overview \cite{jacovi2021formalizing}, there are numerous types of trust that need to be more precisely defined and differentiated. For example, the confusion between two similar, yet different concepts, appropriate trust and appropriate reliance, which often stems from a lack of clear understanding of these terms' definitions. 
Various strategies have been employed to establish an appropriate level of trust in human-AI interaction. Researchers from diverse scientific fields have conducted empirical studies and developed theoretical models to explore different methodologies for building such trust. However, despite the crucial role of appropriate trust in ensuring the successful use of AI systems, there is currently a fragmented overview of its understanding \cite{mehrotratiis}.

To highlight and better understand appropriate trust in human-AI interaction, our paper aims to present a comprehensive overview of the current state of research on Human-AI trust by emphasizing definitions, measures, and methods of fostering \textit{appropriate trust} in AI systems. Furthermore, we make an attempt to map different terms associated with appropriate trust and provide a comprehensive summary of current trends, challenges and recommendations.

\noindent In this work, we study the state-of-the-art in building appropriate trust by examining its evolution, definitions, related concepts, measures, and methods. Our research questions are: 
\begin{enumerate}
    \item What's the history of appropriate trust in automation before AI systems?
    \item How does current research define appropriate trust and what related concepts exist?
    \item How can we structurally make sense of these concepts related to appropriate trust?
    \item What's state-of-the-art in fostering appropriate trust in AI systems? \textit{which includes}
    \begin{enumerate}
        \item How do studies measure whether the trust is appropriate or not?
        \item What kind of tasks do researchers employ in their studies to understand appropriate trust? 
        \item What different types of methods for building appropriate trust exist?
        \item What are the results of the methods aimed at building appropriate trust?
    \end{enumerate}
\end{enumerate}

\noindent To investigate the questions above, we provide a history of appropriate trust development and present a systematic review to identify current practices in the theoretical and experimental approaches. Furthermore, we identify potential challenges and open questions, allowing us to draw research opportunities to understand appropriate trust. First, we provide an overview of the history of understanding appropriate trust in automated systems. Next, we describe our systematic review methodology and the corpus, summarize the current understanding of appropriate trust and propose a Belief, Intentions, and Actions (BIA) mapping to highlight commonalities and differences between concepts. Following this mapping, we present the results of the systematic review, discussing different ways to measure appropriate trust, types of tasks used, approaches to building it, and results of the appropriate trust interventions. Finally, we discuss the challenges identified in studying appropriate trust and summarize our observations as current trends, potential gaps, and research opportunities for future work.


Our main contributions are:
\begin{enumerate}
   \item A Belief, Intentions, and Actions (BIA) based mapping of appropriate trust and related concepts.\\ \textit{Our mapping is result of analyzing how authors define and quantify the abstract notion of appropriate trust and related concepts such as warranted trust, justified trust and meaningful trust};
   \item An exhaustive presentation of different definitions used, measures of appropriate trust, tasks adopted by authors and various methods for building appropriate trust.\\ \textit{Our presentation is based on similarities and differences in the approaches that authors have used to define, measure and build appropriate trust in variety of tasks.};
   \item A set of future research opportunities highlighting current trends, challenges and recommendations for future work.\\ \textit{Our set of future research directions results from a structured summary of our analysis based on the implications of the approaches (definitions, methods, tasks and measures) adopted by the authors to foster appropriate trust in human-AI interaction.}  
\end{enumerate}

\section{Background and History of Appropriate Trust}
The topic of appropriate trust has been maturing for years. As technology evolved from automated machines to decision aids, virtual avatars, robots, and finally, AI teammates, appropriate trust has been studied in both depth and breadth across a variety of domains. As discussions of the failures of under- and over-trust in automation begin to appear, researchers started to study how they could calibrate human trust in automation. One of these early studies defined trust calibration as the relation between user reliance and system reliability \cite{bauhs1994knowing}. Trust calibration was studied by looking at how usage of a system over time changed trust levels, \textit{calibrating} it to the demonstrated reliability of the system. The coining of this term marked the beginning of appropriate trust research within computer science communities, influenced by, but distinct from, previous trust research in e.g. psychology and philosophy. 

\begin{figure}
\centering
\begin{subfigure}{.75\textwidth}
  \centering
  \includegraphics[width=\linewidth]{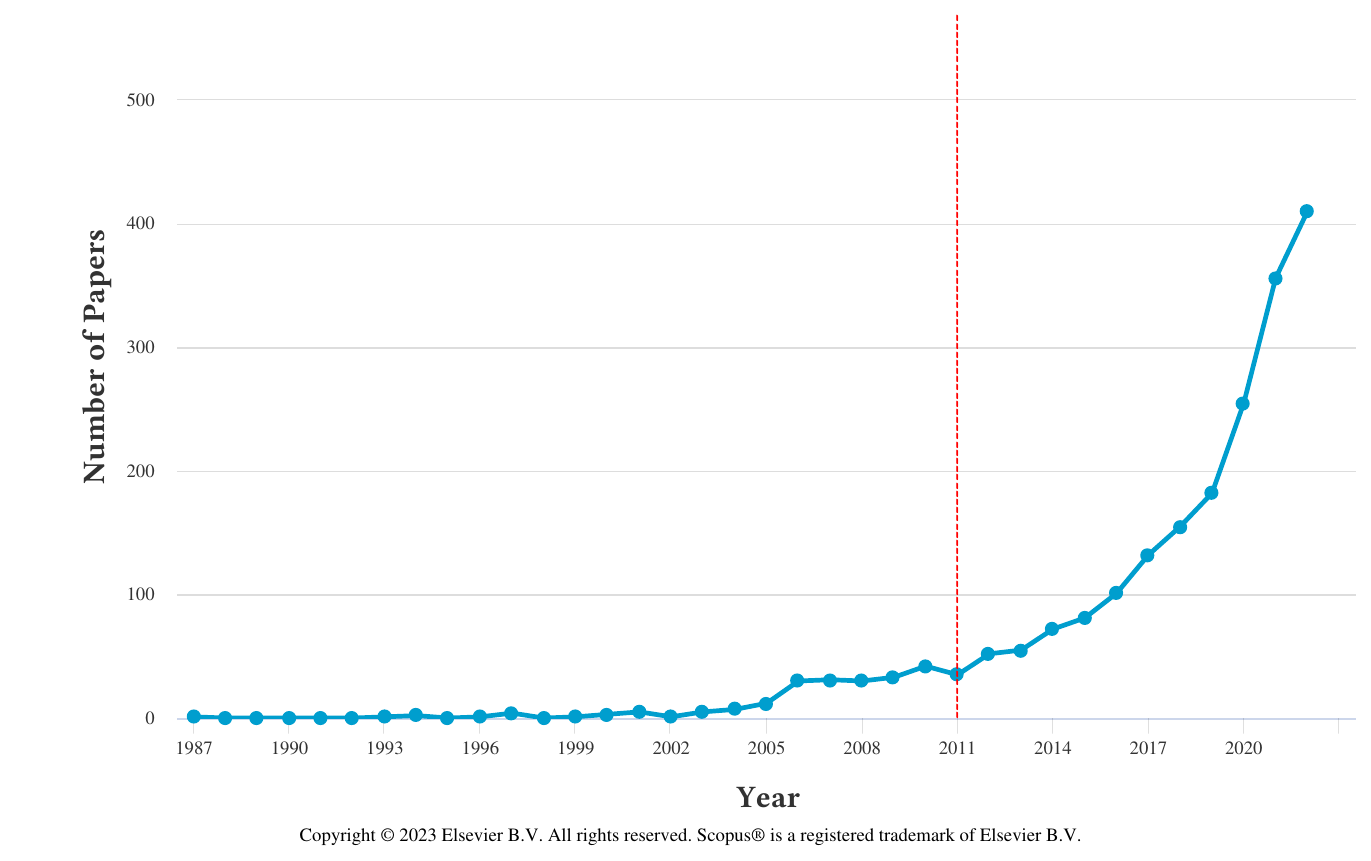}
  \caption{A time-series chart shows the number of studies from 1987 to 2022\footnote{Our search string for this chart was: ( "appropriate trust"  OR  "calibrated trust"  OR  "trust calibration"  OR  "over trust"  OR  "under trust"  OR  "over-trust"  OR  "under-trust" )  AND  PUBYEAR  >  1979  AND  ( LIMIT-TO ( LANGUAGE ,  "English" ) ) }.}
  \label{fig:sub1}
\end{subfigure}%
\begin{subfigure}{.25\textwidth}
  \centering
  \includegraphics[width=1.0\linewidth]{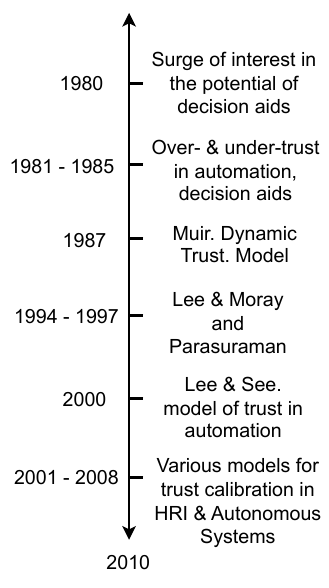}
  \caption{Some key papers during the development process of appropriate trust research.}
  \label{fig:sub2}
\end{subfigure}
\caption{(a) A timeline for the development of appropriate trust as a topic of research from 1987 to 2022 based on the hits from the SCOPUS database. The red dotted line indicates rapid rise of research interest in appropriate trust research. (b) Some key papers during the early stage development of the topic. It's important to note that these are just some key developments and trends in the study of appropriate trust during this time period, and the field has continued to evolve and expand in the years since 2010.}
\label{fig:timeline}
\end{figure}

Understanding the historical context and evolution of appropriate trust allows us to position this work within the broader context of the field. Therefore, in this section, we chronologically describe past efforts to study appropriate trust until the starting point of our systematic search. The background and history of appropriate trust can provide insights about its conceptualization and how technological and social factors have influenced the research field. Moreover, historical analysis can highlight the various theoretical frameworks that have been used to study trust calibration and their limitations. By examining the historical development of this topic, we can better understand its current conceptualization and identify gaps in the literature. In Figure \ref{fig:sub1} \& \ref{fig:sub2}, we illustrate the timeline for these developments.



\subsection{1980-1990s: Over- and under-trust in automation}
The question of how and when to trust automation easily pre-dates the modern computer era. In the early 1980s, there was a surge of interest in the potential of computer-based decision aids to support decision-making in various fields \cite{teach1981analysis,steeb1981computer,kramer1995trust}. As automation gained further computing power and was able to solve tasks with high complexity, people started relying on the advice provided by these systems. However, early studies found that users tended to over-trust this advice, even in cases where it was clearly incorrect or irrelevant \cite{hurst1982pilot,sheridan1987handbook}. This phenomenon was referred to as "automation bias" or "automation-induced complacency" \cite{wiener1981complacency}. This concern populated further in the late 1980s, where researchers were concerned about the reliability and safety of nuclear power plants. 

Over-trust in automation is only one side of the coin, while under-trust is the other. One of the factors identified as contributing to various accidents such as the Baltimore train incident \cite{10.1121/1.397232} or misuse of anti-ballistic missiles \cite{feigenbaum_1971} was the tendency of operators to under-trust the information provided by the control systems and not to rely on them. This problem led to the development of various training and simulation programs aimed at improving operator trust in the automation \cite{bainbridge1983ironies}.

In this era, researchers were interested in understanding how humans interact with automated systems and errors that can occur when trust is misplaced. Studying the operator role and human-system integration, Knee and Schryver found that over- and under-trust stem partially from consistent, reliable performance by the Intelligent Machines (IM) within tasks, problems, etc. that the human operator may not fully understand (due to the lack of training, experience, or even the ability to be actively involved in system operation) \cite{knee1989operator}. According to them, such cases may support "blind reliance" on the part of the human operator, i.e., acceptance of IM control actions without question of its intent or motives. In conclusion, the study of trust in automation from the 1980s to the 1990s sheds light on the pitfalls of misuse, disuse, and overuse of automated systems, highlighting the importance of understanding how humans trust automated systems. 

\subsection{1990s: Introduction of HCI as a field and focus on appropriate trust}
In 1987, Muir presented a model based on dynamics of trust between humans and machines for calibrating user's trust in decision aids \cite{MUIR1987527}. 
At this point in time, extensive research began in the Human-Computer Interaction (HCI) community to examine the factors that influence a human’s trust in automation \cite{hoff2015trust}. 
One of the themes of this research was calibrated trust. 

In the CHI '94 conference, Bauhs and Cooke showcased the effect of system information on trust calibration \cite{bauhs1994knowing}. The authors reported that the system information aided in calibrating users' confidence in system reliability, but it had little effect on users' willingness to take expert system advice. In the same year, Lee \& Moray showed how trust and self-confidence relate to the use of automation and refereed trust calibration as correspondence between a person's trust in automation and the automation's capability \cite{lee1994trust}. Following Lee \& Moray's work, a seminal article by Parasuraman and Riley \cite{parasuraman1997humans} in 1997 on the use, disuse, abuse, and misuse of automation indicated the issue of over- and under-reliance on machines due to lack of trust. 

Many articles followed Lee \& Moray and Parasuraman \& Riley research. Ostrom in 1998 \cite{ostrom1998behavioral} showcased that effectively studying trust in automation can help alleviate the uncertainty in gauging the responses of others, thereby guiding appropriate reliance. Tangentially, Kaber, and Endsley introduced the concept of situational awareness to tackle the issue of mistrust in automated systems in the same year \cite{endsley1999SA}. Thus, the emergence of trust calibration studies signalled and ushered in a greater focus on user-centered design as a means of minimizing automation dis- and mis-use. 

\subsection{2000s: Emergence of appropriate trust as a key topic of research}
A seminal article by Lee \& See in 2004 provided the first conceptual model of the relationship among calibration, resolution, and automation capability in defining appropriate trust in automation \cite{lee2004trust}. This work by Lee \& See was built on the key work by Cohen et al. in 1998 \cite{cohen1998}. The Lee \& See model was based on purpose, process, and performance dimensions of information that describe the goal-oriented characteristics of the agent to maintain an appropriate level of trust. 

In 2006, Duez et al. \cite{duez2006trust} followed Lee \& See's model to study information requirements for appropriate trust in automation, while Dongen and Maanen \cite{van2006under} investigated whether calibration improves after practice and whether calibration of own reliability differs from calibration of the aid's reliability. Thus, researchers were able to develop models of information communication \cite{duez2006trust} and asymmetrical reliability attribution \cite{van2006under} in automated systems, which improved understanding of how users calibrated their trust over time. 
Following the mentioned works and literature on calibrated trust, the Human-Robot Interaction community developed an of understanding appropriate trust in robot capabilities, such as Freedy's et al. measures of trust in human-robot interaction for detecting over- and under-trust in 2007 or Hancock \textit{et al.}'s 2011 meta-analysis of factors affecting trust in Human-Robot Interaction \cite{hancock2011meta}. Their results indicated that improper trust calibration could be mitigated by manipulating robot design, focusing on quantitative estimates of human, robot, and environmental factors. Similarly, Sanders et al. \cite{sanders2011model} provided a model of human-robot trust targeting performance, compliance, collaboration, and individual human differences to study how human trusts can be calibrated in situations of over- and under-reliance.

The topic of appropriate trust also started to pick up in industrial settings during the 2000s. For example, in 2008, Wang and their colleague from a defense R\&D studied the effectiveness of providing aid reliability information to support participants' appropriate trust in and reliance on a combat identification aid \cite{wang2008improving}. Their results showed that participants who needed to be made aware of the aid's reliability trusted in and relied on the aid feedback less than those who were aware of its reliability, highlighting appropriate reliance on the aid. 

Thus, the emergence of appropriate trust as a prominent topic in the 2000s was marked by the increasing prevalence of automation and innovative steps taken by researchers to study the role of this topic. Notably, Lee \& See's 2004 article \cite{lee2004trust} introduced a conceptual model that interconnected calibration, resolution, and automation capability to define appropriate trust in automation. This work which was built on Cohen's et al. work \cite{cohen1998} was followed by many authors such as \cite{duez2006trust,hancock2011meta,sanders2011model,wang2008improving} where fresh insights were seen considering purpose, process, and performance dimensions of information, offering a deeper understanding for trust calibration. Furthermore, the relevance of appropriate trust extended to industrial settings, as demonstrated by studies on combat identification aids and defense technology \cite{wang2009trust}. 

\subsection{Parallel Developments: Influential Domains}
While research in automation has made significant contributions to our understanding of how people develop and calibrate their trust in computer systems, appropriate trust is also studied in a variety of other fields, including psychology and philosophy. In many cases, our current understanding of appropriate trust have in fact stemmed from the paradigms established in these domains \cite{lee2004trust,freedy2007measurement,parasuraman2004trust}. 

Different disciplines study appropriate trust differently, however they all seek the capacity for accurate trust assessment, with the goal of establishing a robust basis for augmented decision-making. Appropriate trust has been studied extensively in \textit{psychology}. It is understood as trust that is based on a rational assessment of the risks and benefits of trusting another person or source of information \cite{schaubroeck2011cognition,lewicki1996developing}. For example, Evans et al. showcased that older children (9-10 years) are more sensitive to changes in trustee's characteristics, suggesting that they are not only more trusting, but more discerning in their decisions of when to trust \cite{evans2013development}. Similarly, Barnard showcased that how medical professionals change their attitudes and behaviors to gain trust of their patients and proposed which conditions would win justified trust\footnote{Different terms have been used in the literature which are related to appropriate trust such as "Justified Trust", "Optimal Trust", etc., refer Section 5.1 for details.} of patients in them \cite{barnard2016vulnerability}. Therefore, in \textit{human-human interaction} the ability to accurately calibrate trust is essential for building and maintaining strong relationships, as it helps individuals to avoid betrayals and to cultivate mutual respect and understanding. Overall, appropriate trust is an important aspect of social functioning and well-being. 

In \textit{Philosophy}, the concept of appropriate trust is closely related to the idea of epistemic responsibility, which emphasizes the need for individuals to take responsibility for their beliefs and to use appropriate methods for evaluating evidence and making judgments \cite{rousseau1998not,pouryousefi2022empirical,beauchamp1995moral}. In particular, according to Onora O'Neill, appropriate trust involves a "reasonable reliance on another's goodwill, competence, and reliability" \cite{o2002autonomy}. Other philosophers, such as Karen Jones \cite{jones2018politics} and Katherine Hawley \cite{hawley2017trustworthy}, have also explored the concept of appropriate trust and the importance of carefully calibrating one's trust based on various factors, such as past experiences, social norms, and situational factors. 


The research interest in trust calibration evolved slowly compared to promoting trust in automation \cite{cho2015survey}. This can be partly due to a higher interest in understanding multidimensional aspects of trust, and partly due to the complex nature of automation systems. However, in the last ten years (2012-2022), interest in appropriate trust research has grown drastically, see Figure: \ref{fig:sub1}. This trend is likely driven by the increasing cognitive complexity of AI and ubiquity of interpersonal interactions, as well as organizational interest. Therefore, it has become timely to provide an in-depth literature overview of the state-of-the-art for building appropriate trust in AI. We follow the methodology outlined in this section to provide a comprehensive overview of studies from 2012 till June 2022 in the following section with our systematic review methodology.

\section{SYSTEMATIC REVIEW METHODOLOGY}
\begin{figure}[ht!]
    \centering
\includegraphics{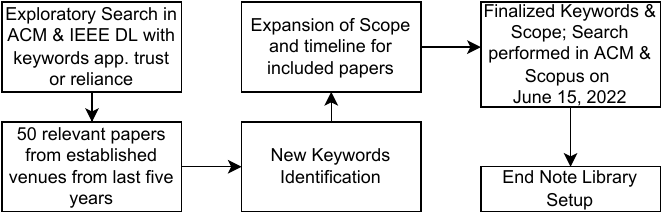}
    \caption{Search process for preparing the corpus of the systematic review}
    \label{fig:paper search}
\end{figure}
\begin{figure}[h]
    \centering
\includegraphics{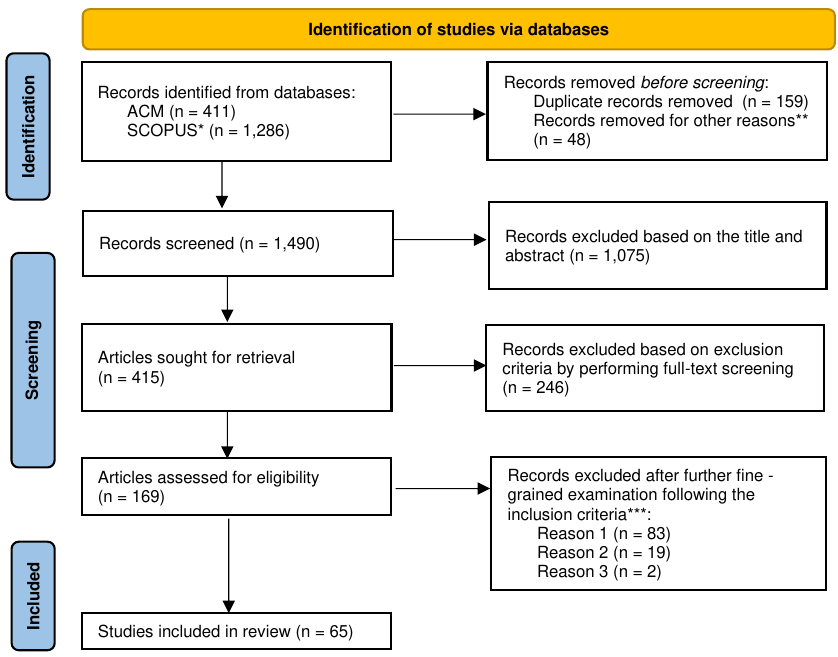}
    \caption{*SCOPUS data includes all retrieved databases such as IEEE, Springer, SAGE, etc. ACM data was excluded from the search as it was taken from ACM DL.\\
** Other reasons include: records not retrieved, broken URL, and blank pages in the published record.\\
*** Reason 1: The article’s focus is NOT on appropriate trust derived from the primary or secondary research question. Reason 2: The article does not use a method or a measurement technique to measure or calculate trust. Reason 3: The article is published as a short version of a long paper (In this case, we included the longer version of the article).
}
    \label{fig:prisma}
\end{figure}

\noindent We conducted a systematic review to understand (a) current understanding about building appropriate trust in AI, (b) how appropriate trust and its related concepts have been defined and conceptualized, and (b) what measures and methods have been made to achieve appropriate trust in AI. We adapted the procedure by Calvaresi et al. \cite{calvaresi2018multi}  by developing the research protocol following inclusion and exclusion criteria. For search and identification of the relevant articles, we followed the PRISMA guidelines \cite{page2021prisma}. The specifications of these guidelines is illustrated in Figure \ref{fig:prisma}.

\subsection{Search String}\label{search_string}
Appropriate trust is a complex concept and the term 'appropriate' is often interchangeably used with terms for similar concepts (such as \textit{appropriate reliance, justified trust}, etc.) \cite{jacovi2021formalizing,tolmeijer2022amoral}. Therefore, we first conducted an exploratory search to determine which terms for similar concepts are used. In the ACM and IEEE Digital Libraries, we searched for articles with the keywords ``appropriate trust" or ``calibrated trust" from the last five years\footnote{This phase was conducted in May 2022. We decided for the last five years as it coincides with the recent rise of interest in appropriate trust research.}. This exploratory search produced 186 results. Among these 186 results, we focused on articles from four of the most most reoccurring and relevant computer science venues, FAccT, CHI, IUI, and HRI. We selected 50 articles (FAcct: 6, CHI: 20, IUI: 12, and HRI: 12) with the highest use of keywords and similar concepts throughout the articles. 

We manually reviewed every title, keyword, and abstract to find new keywords to be included in our final search string (e.g., ``optimal trust", ``justified trust"). We iterated different combinations of the keywords until all papers deemed relevant in the exploratory step appeared among the ACM \& IEEE Digital Libraries search results. Analyzing the text of the relevant articles and their references, we noticed that scholars from the Computer Science community often cite scholars from other disciplines who also study appropriate trust. These disciplines include engineering, social sciences, psychology, mathematics, and decision sciences. Therefore, we decided to include these subjects in our search criteria. Furthermore, we decided to broaden our timeline to include articles published in the last decade\footnote{Since the aim is to identify the current trends and understand recent works in appropriate trust research, we chose to restrain this work to papers published in the last decade} (2012-2022) after examining the references of the articles. Figure \ref{fig:paper search} visualizes our search process and string finalization. The final search string used in ACM and SCOPUS search is:
\begin{verbatim}
( ( "appropriate trust" )  OR  ( "calibrated trust" )  OR  ( "warranted trust" )  OR  ( "justified trust") 
OR  ( "optimal trust" )  OR  ( "responsible trust" )  OR  ( "trust calibration" )  OR  ( "over trust" )  OR  
( "under trust" )  OR  ( "over-trust" )  OR  ( "under-trust" )  OR  ( "meaningful trust" ) )  AND  PUBYEAR  
> 2011  AND  PUBYEAR  <  JUL 2022  AND  ( LIMIT-TO (SUBJAREA ,  "COMP" )  OR  LIMIT-TO ( SUBJAREA ,  "ENGI" )
OR  LIMIT-TO ( SUBJAREA ,  "SOCI" )  OR LIMIT-TO ( SUBJAREA ,  "PSYC" )  OR  LIMIT-TO ( SUBJAREA ,  "MATH" )
OR  LIMIT-TO ( SUBJAREA , "DECI" ) )  AND  ( LIMIT-TO ( LANGUAGE ,  "English" ) )    
\end{verbatim}

\subsection{Selection Criteria}
Our search string generated 1,697 articles from the ACM and SCOPUS databases. This phase of generating the final list of articles based on the search string was conducted on June 15, 2022. The screening of articles was carried out manually in three stages: (A) title and abstract screening based on the inclusion criteria, (B) full-text screening based on the exclusion criteria, and then (C) full-text screening with a fine-grained examination based on the inclusion criteria. Our inclusion criteria were:
\begin{enumerate}
    \item \textbf{Language}: The article should be in English.
    \item \textbf{Peer-Reviewed}: The article should have been peer-reviewed. For example, articles from arxiv, OSF, magazine articles, etc., were excluded.
    \item \textbf{Format}: Only full and short articles were included so that all the reviewed articles could contain similar details about a study. Therefore, posters, dissertations, workshop papers, workshop calls, etc., were excluded.
    \item \textbf{Publication Singularity}: Only the complete version of the article is included. 
    \item \textbf{Human-Centered}: Studies needed to have some form of human involvement to be included. For instance, full simulated multi-agent models were excluded. 
    \item \textbf{Inclusion of a Definition}: For a paper to be included, it should have a explicit definition or implicit definition through either referencing previous work or describing measures of appropriate trust or the similar concept (calibrated trust, warranted trust, etc.). 
     \item \textbf{Conceptualization of Appropriate Trust}: The articles should conceptualize appropriate trust with measurable constructs or a similar concept. For example, the article uses measurable constructs for appropriate trust.
\end{enumerate}

After applying the inclusion criteria in a two-step abstract and full-text screening process, 169 articles remained. On these 169 articles, we performed further fine-grained examination based on the following criteria:

\begin{enumerate}
    \item \textbf{Contribution Scope}: Articles whose primary contribution was unrelated to appropriate trust were excluded. Articles discussing the need for appropriate trust without any direct contribution to define, measure, or model it were also excluded.
    \item \textbf{Contribution Type}: Surveys, Scoping Reviews, and Literature Reviews were excluded.
\end{enumerate}

The research team registered the protocol of the review with Open Science Foundation (OSF)\footnote{\url{https://osf.io/c78tw/?view_only=16c398038f474b9b8922277a3fd94c87}} to make the selection of reviews a transparent process. Once the registration was completed, the first and second author independently examined the full text of 169 articles. Both authors used the Rayyan web app \cite{ouzzani2016rayyan} to organize their decisions. 
When there were discrepancies between their decisions, the two researchers involved the senior author in discussing it. This discussion process resulted in the final list of 65 articles for the systematic review.

\subsection{Corpus Overview and Analysis}
\begin{figure}[h!]
     \centering
     \begin{subfigure}[b]{0.5\textwidth}
         \centering
         \includegraphics[width=\textwidth]{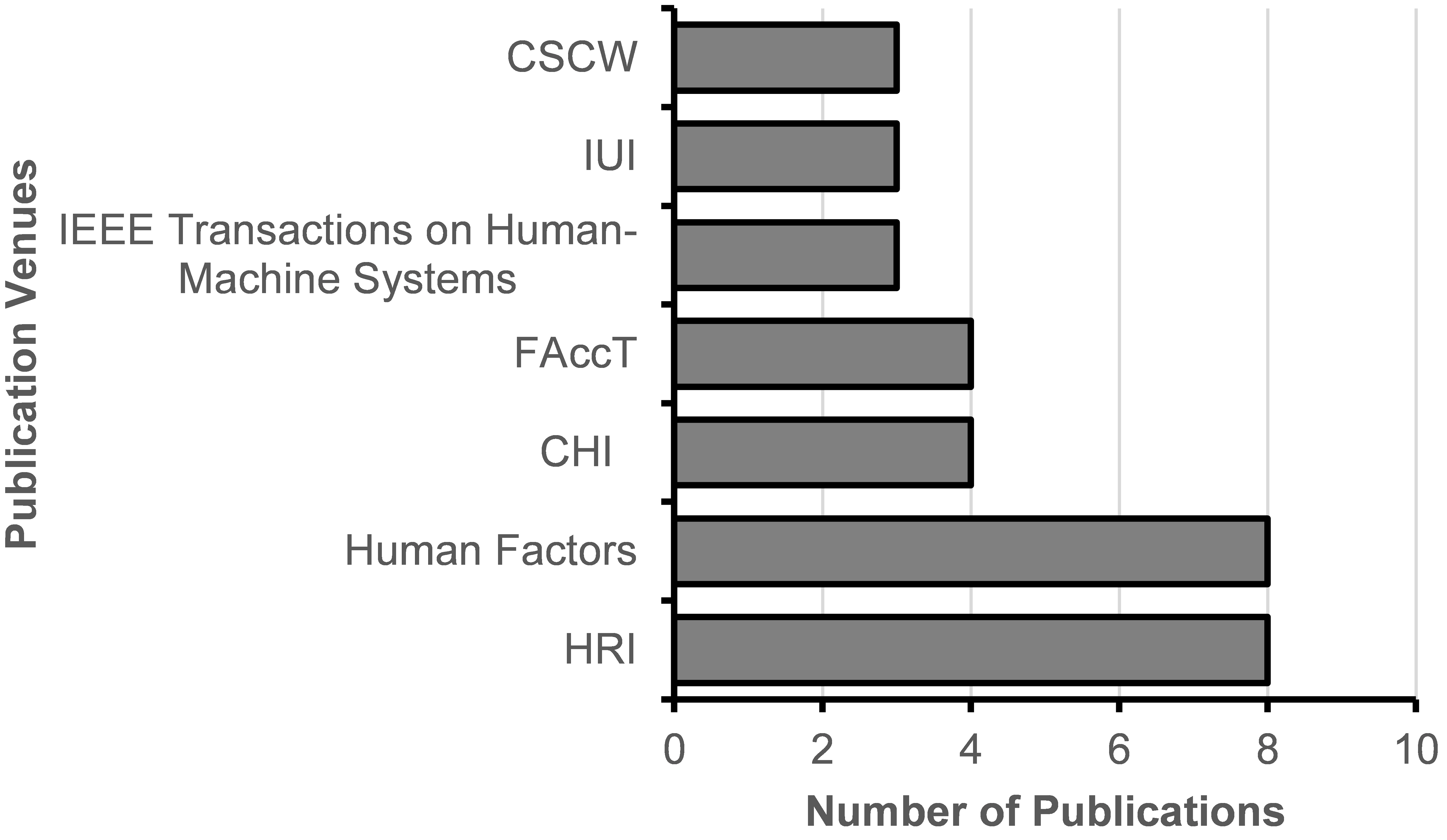}
         \caption{Number of the selected papers per top six publishing venues}
         \label{fig:publications}
     \end{subfigure}
     \begin{subfigure}[b]{0.49\textwidth}
         \centering
         \includegraphics[width=\textwidth]{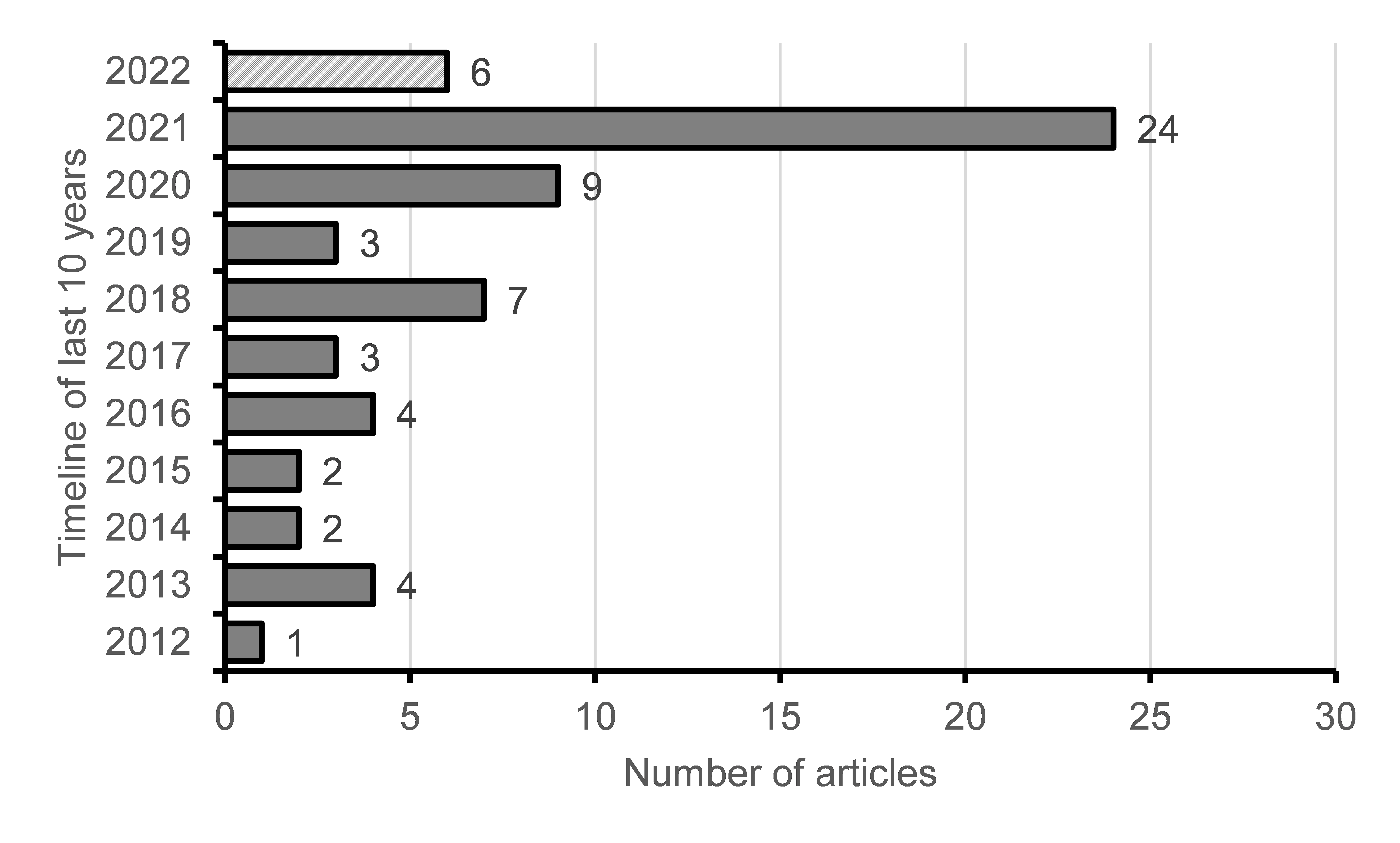}
         \caption{Number of papers per year from 2012 to 2022}
         \label{fig:pub_timeline}
     \end{subfigure}
        \caption{Distribution of selected articles over last ten years and top six of their publication venues. Please note that the data for 2022 is incomplete since the data collection for this literature review was conducted in the mid of June 2022.}
        \label{fig:two graphs}
\end{figure}

The final corpus consists of 65 papers, on which we first performed a metadata analysis. We were interested in the publication venues, timeline of publications, and application scenarios. The top six publication venues and chronological distribution of articles over the last ten years are shown in Figure \ref{fig:two graphs}a and \ref{fig:two graphs}b.

In Figure \ref{fig:publications}, we can observe that the most popular venues, among others are Human Factors and Ergonomics Society (Human Factors) and HRI (n = 8 each) and CHI \& FAccT (n = 4 each, idem) which account for 47.3\% of the final corpus. 
Also, the last five years have experienced a growth in the number of publications related to the appropriate trust, see Figure \ref{fig:timeline}. 
This trend reflects a growing interest in human-centered AI and the importance of studies focusing on appropriate trust, distinct from enhancing trust in AI. 
A spreadsheet containing list of all final papers can be found with the pre-registration link on OSF.

\section{Definitions and Related Concepts}
Appropriate Trust in AI systems is growing rapidly as a research field for both researchers and practitioners. To understand how to achieve appropriate human trust in AI (Human-AI trust), it is important to understand how we define it and its related concepts. 
The increasing interest in Trustworthy AI research \cite{thiebes2021trustworthy} has brought to light a growing need for clarity among the community regarding the different concepts and definitions related to appropriate trust in AI. 

Terms like “appropriate trust”, “calibrated trust” and "appropriate reliance" are often used interchangeably in prior research \cite{schemmer}. There have been debates in the community about what appropriate trust is and how different concepts related to appropriate trust are different or similar, for instance during the CHI TRAIT workshop in 2022 \cite{trait} and the CSCW '23 workshop on "Building Appropriate Trust in Human-AI Interactions" \cite{alizadeh2022building}. These debates are a result of the complex nature of trust in AI systems, which can be difficult to understand and evaluate. In this systematic review, we identified different terms related to appropriate trust in the literature, the most common ones being calibrated trust (number of articles (n) =16), appropriate reliance (n=8) and warranted trust (n=6). The full list of terms is available in Table \ref{tab:definitions}, with the corresponding definitions as given by the papers. We can see from Table \ref{tab:definitions} that there is often more than one definition of appropriate trust or its related concepts. This discord and diversity among different concepts motivates us to establish links between them and present a unified mapping.

\begin{table*}[htbp]
  \caption{Definitions of Appropriate Trust and its related concepts, A * represents articles before the year 2012 or after the end of search date. The * articles were not included in the review process.}
  \label{tab:definitions}
  \begin{tabular}{p{0.15\linewidth} | p{0.8\linewidth}}
    \toprule
    Keyword &Definition \\
    \midrule    
   {\footnotesize\texttt{\textbf{Appropriate Trust} - Based on System Performance or Reliability:
    }}  & {\footnotesize\textit{
1. Appropriate trust is the \textbf{alignment between the perceived and actual performance} of the system. Appropriate trust is to [not] follow an [in]correct recommendation. Other cases lead to over-trust or under-trust"  \cite{yang2020visual}.\newline
2. If the \textbf{reliability of the agent} matches with user’s trust in the agent then trust is appropriately calibrated \cite{okamura2020adaptive}.\newline
3. In human-robot teaming, appropriate trust is maintained when the human uses the robot for tasks or subtasks the robot \textbf{performs better} or safer while reserving those aspects of the task the robot \textbf{performs poorly} to the human operator \cite{ososky2013building}. }} \\ \cline{2-2}
{\footnotesize\texttt{Based on TW and beliefs:}}&
{\footnotesize\textit{1. Appropriate trust in teams happens when one teammate’s trust towards another teammate corresponds to the latter’s
\textbf{actual trustworthiness} \cite{jorge2021trust}.\newline 
2. We can understand ‘appropriate trust’ as obtaining when the trustor has \textbf{justified beliefs} that the trustee has suitable dispositions \cite{danks}.}}\\
\cline{2-2}
{\footnotesize\texttt{Based on the Calculations:}}&
{\footnotesize\textit{1. “Appropriate trust is the \textbf{fraction of tasks} where participants used the model’s prediction when the model was
correct and did not use the model’s prediction when the model was wrong; this is effectively participants’ final
decision accuracy"  \cite{wang2021explanations}.\newline 
2. FORTNIoT (a smart home application) predictions lead to a more appropriate trust in the smart home behavior.
Meaning, we expect participants to \textbf{have reduced under-trust} (i.e. they trust the system more when it is behaving
correctly) \textbf{and reduced over-trust} (i.e. they trust the system less when it is behaving incorrectly) \cite{coppers2020fortniot}.\newline 
3. \textbf{Trust appropriateness was calculated} by subtracting a{\_}ideal from a participant’s allocation for a given round.
Thus, a positive value indicates trust that is too high, a negative value indicates trust that is too low, and 0 indicates calibrated, appropriate trust \cite{jensen2021trust}. \newline
4. The level of trust a human has in an agent with respect to a contract is appropriate if the likelihood the human associates with the system satisfying the contract is equal to the \textbf{likelihood} of the agent satisfying that contract* \cite{zahedi2023mental}. \newline
5. The term appropriate trust then is the \textbf{sum of appropriate agreement and appropriate disagreement} of humans with the AI prediction\cite{liu2021interactiveexplain}}}\\ \midrule
{\footnotesize\texttt{\textbf{Warranted Trust}}}& {\footnotesize\textit{1. “Warranted trust describes a match between the actual \textbf{system capabilities} and those perceived by the user" \cite{schlicker2021towards}.\newline
2. “Human’s trust in a AI model (to Contract - C) is warranted if it is caused by \textbf{trustworthiness in the AI model}. This holds if it is theoretically possible to manipulate AI model’s capability to maintain C, such that Human’s trust in AI model will change. Otherwise, Human’s trust in AI model is unwarranted." \cite{jacovi2021formalizing}}}\\\midrule
{\footnotesize\texttt{\textbf{Justified Trust}}}& {\footnotesize\textit{1. “Justified Trust is computed by evaluating the human’s understanding of the model’s decision-making process.
In other words, given an image, justified trust means users could \textbf{reliably predict} the model’s output decision."
\cite{akula2020cocox} }}\\ \midrule
{\footnotesize\texttt{\textbf{Contractual Trust}}}& {\footnotesize\textit{1. “Contractual trust is when a trustor has a belief that the trustee will stick to a \textbf{specific contract}". \cite{jacovi2021formalizing}\newline
2. “Contractual trust is a \textbf{belief in the trustworthiness} (with respect to a contract) of an AI." \cite{ferrario2022explaintrust}}}\\\midrule
{\footnotesize\texttt{\textbf{Calibrated Trust}}}& {\footnotesize\textit{1. “Trust calibration is the \textbf{process by which a human adjusts their expectations} of the automation’s reliability and trustworthiness". \cite{lebiere2021adaptive}\newline
2.  Calibrating trust is if explanations could help the annotator \textbf{appropriately increase their trust} and confidence as
the model learns \cite{ghai2021explainable}\newline
3. Trust calibration refers to the correspondence between a person’s trust in the automation and the automation’s capabilities* (\textit{based on Lee \& Moray \cite{lee1994trust} and Muir \cite{MUIR1987527})} \cite{lee2004trust}. }}          \\\midrule
{\footnotesize\texttt{\textbf{Well-placed Trust*}}}& {\footnotesize\textit{“[T]he only trust that is \textbf{well placed} (intention) is given by those who understand \textbf{what is proposed}, and who are in a position to refuse or choose in the light of that understanding \cite{o2002autonomy}. }}  \\\midrule
{\footnotesize\texttt{\textbf{Responsible Trust}}}& {\footnotesize\textit{“The area for responsible trust in AI is to explore means to \textbf{empower end users to make more accurate trust judgments}". \cite{liao2022designing}. }} \\\midrule
{\footnotesize\texttt{\textbf{Reasonable Trust*}}}& {\footnotesize\textit{“Reasonable
trust requires \textbf{good grounds for confidence in another’s good will}, or at least the absence of good grounds for expecting their ill will or indifference.". \cite{baier1986trust} }}\\
\bottomrule
\end{tabular}
\end{table*}

\subsection{A Belief-Intentions-Actions (BIA) Mapping}
Given the number of terms and slightly different definitions that exist, our first aim is to achieve a clearer understanding of the different concepts surrounding appropriate trust. To this end, we grouped all the presented concepts at different levels of human perception in a conceptual mapping, following Michael Bratman's theory of human-practical reasoning \cite{Bratman1987-BRAIPA}. In this subsection, we will first discuss the relationships among appropriate trust and its related concepts following this mapping. Following, we relate the concepts to the definitions presented by the authors of the included papers.

\begin{figure}[h]
    \centering
    \includegraphics[scale=0.8]{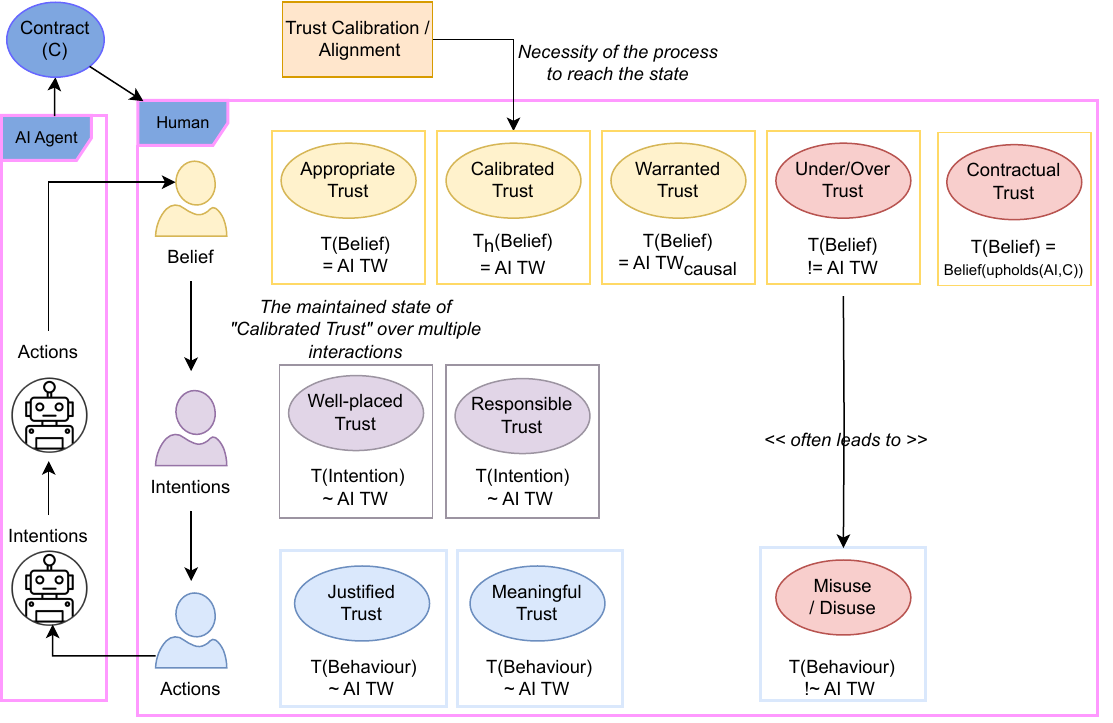}
    \caption{In this figure, we present a Belief, Intentions and Actions (BIA) mapping of Appropriate Trust and related concepts. The pink outlines represent the elements linked with the human and the AI agent entity. The black coloured circle with a robot icon represents the AI agent. For brevity, when writing T(Belief), we mean T(human,agent) (Belief) and for TW(agent) we use AI TW. Also, Under/Over trust and Contractual trust are represented in different colors as these types of trust aren't (necessarily) appropriate.}
    \label{fig:trust_concept_map}
\end{figure}

We illustrate our categorization of the concepts associated with appropriate trust in Figure \ref{fig:trust_concept_map}. Figure \ref{fig:trust_concept_map} presents a Belief, Intentions, and Actions (BIA) mapping of appropriate trust and related concepts. These levels allow us to separate the different perspectives on trust as a belief, intention, or action. More specifically, \textit{Beliefs} describe a perception of the world and the other agents in it, including beliefs about other agents' intentions and actions. Beliefs may or may not be justified based on current information about the world and past agent behavior \cite{ferrario2022explaintrust}. Second, \textit{Intentions} represent the deliberative state of the human – what the human has chosen to do. Intentions are desires to which the human has to some extent committed \cite{georgeff1999belief}. Finally, \textit{Actions} describe events as they actually occur in the interaction \cite{allen1984towards}, such as a doctor actually offering a patient an in-person consultation. In essence, this mapping provides a mechanism for separating each interaction into three parts; the informational state (beliefs), motivational and deliberative states (intention), and reactive activity (actions).

Our mapping identifies two actors: the Human and the AI agent. The human actor is illustrated with a `user' icon and the AI agent is represented by a robot icon. To help distinguish between the different concepts, we formally define them. 
In our definitions, we use the following variables: $human\in H$ for a human, $agent \in A$ for an AI agent, and $T_{(x,y)}$ to denote the trust that trustor x has in trustee y.  
We then use the following notations:
\[\textit{T$_{(human, agent)}$}(Belief) = \text{Trust of the human in the AI agent is about the human's belief about the agent}\]
\[\textit{TW$_{agent}$} = \text{Actual trustworthiness of the agent}\]
\[\textit{T$_{(human, agent)}$}(Intention) = \text{Trust of the human in the agent is about the human's intentions towards the agent}\]
\[\textit{T$_{(human, agent)}$}(Behaviour) = \text{Trust of the human in the agent is about the human's behaviour towards the agent}\]

We have divided the definitions of appropriate trust in the Table\ref{tab:definitions}, see keyword "Appropriate Trust" based on the similarities such as system performance or reliability, beliefs, and calculations. Based on this division, we can formulate our conceptualization of appropriate trust. We define trust to be appropriate when the human's trust formed by beliefs about the AI agent's trustworthiness (denoted as \textit{TW$_{(human, agent)}$(Belief)}) is equal to the AI agent's actual trustworthiness \textit{TW$_{agent}$(Actual)}, refer eq. \ref{eq:app_trust}.

\begin{equation}
\label{eq:app_trust}
    \text{\textbf{Appropriate Trust}} \iff T_{(human, agent)}(Belief) = \textit{TW$_{Agent}$}
\end{equation}

From this foundation, we go on to differentiate between the many related concepts for appropriate trust which we have encountered. As a note to our readers, the definitions and terms presented in the Table \ref{tab:definitions} don't always match one-to-one with our conceptualization in Figure \ref{fig:trust_concept_map}, because sometimes different authors define the same term in different ways.

First, we consider \textbf{calibrated trust}, the most common term in the reviewed corpus, which introduces notions of dynamic trust and trust variations to the Human-AI interaction \cite{mcbride2010trust,schaefer2016meta}. We define calibrated trust as similar to the appropriate trust in that a human's trust belief about the agent corresponds to their actual trustworthiness. However, calibrated trust necessarily involves a process of \textit{trust calibration} or \textit{trust alignment} that corrects for over- and under-trust over the course of time and repeated interactions. We postulate that appropriate trust is the maintained state of ``calibrated trust" over multiple interactions. Nevertheless, human trust in an AI system may be appropriate even without calibration. 

Distinct from appropriate or calibrated trust is \textbf{over-trust} \textit{i.e.,} the human's trust beliefs in an AI agent's is greater than the AI agent's actual trustworthiness \textit{TW$_{Agent}$}. Similarly, when the human's trust belief in an AI agent is less than the AI agent's actual trustworthiness \textit{TW$_{Agent}$} then a state of \textbf{under-trust} is reached.

Next, \textbf{warranted trust} is defined as trust caused by the trustworthiness of the AI agent. More specifically, we talk about warranted trust if there is a causal relationship between trustworthiness of the trustee and the trust of the trustor \cite{ferrario2022explaintrust}. Though we expect warranted trust to mostly be appropriate, not all appropriate trust needs to be warranted. In other words, while trust that is well-supported by evidence and reasoning is probably appropriate, there may be situations where trust is appropriate even if there is no clear evidence or justification to support it. For instance, if an e-commerce website has a polished and visually appealing design, it may create an initial positive impression in the user's mind. This positive impression, in turn, may lead the user to trust the website's content to some degree, even though they lack in-depth evidence about the product's quality. Finally, \textbf{contractual trust} in the AI agent is based on the belief that the AI agent will uphold an explicit contract (upholds(AI,C)) which specifies what the AI agent is expected to do \cite{hawley2014trust,tallant2017commitment}. Here, the contract may refer to any functionality of the AI agent that is deemed useful, even if it is not concrete performance at the end-task that the AI agent was trained for \cite{jacovi2021formalizing}. It is important to highlight that contractual trust differs from many other definitions in that it does not directly imply appropriateness, as the human's beliefs about the agent might not be related to its actual trustworthiness.

Unlike the three above-mentioned concepts, \textbf{well-placed and responsible trust} are built around intentions. Here, both well-placed and responsible trust are defined as intentions about how to act towards the agent (denoted as \textit{TW$_{human}$(Intent)}). Meaning, if a human has well-placed trust, that means its intentions are correct given the trustee's trustworthiness. 

Trust is justified when a human's behavior is appropriate given the agent's trustworthiness.
In this case, the human actor is acting trustingly towards an agent. 
The ability of a user to evaluate trust does not make the AI agent more accurate, robust, and reliable in itself; it can only, at best, make the use of the AI by the human more accurate, robust, and reliable leading to
\textbf{justified trust}. In contrast, when human’s trust is not justified based on the AI’s agent trustworthiness, it is plausible that the user can lean towards misuse or disuse of AI. 

So far, we have described our mapping related to the human actor. Now, we shift our focus to the AI agent. As mentioned earlier, we consider the AI agent attributed with intent. Here, the AI agent can form an intention based on the human behavior which can be associated with the action(s) it can take. 
Finally, the AI agent can set expectations for a human to form beliefs about the AI agent's trustworthiness, making it a closed-loop process. This process helps in arriving at the belief that a contract, as outlined by Jacovi et al \cite{jacovi2021formalizing}, has been established between the pair and will be adhered to in the future. In other words, an AI agent can form an intention based on human behavior or actions by observing the human behavior. Based on its intention it can decide how to respond by taking an action. By doing so, the AI agent can ensure that it acts in accordance with the contractual agreement (actions the AI agent is authorized to take and the expectations for how the AI agent should behave) and maintains the trust of the human.

In addition to the aforesaid concepts, we found a few further, more minor and less defined terms which are not completely covered in Figure \ref{fig:trust_concept_map} and these are the ones which we haven't defined explicitly, namely meaningful trust (closely linked to \textit{Justified Trust}), optimal trust (\textit{Justified Trust}), moral trust (\textit{Responsible Trust}), capacity trust (\textit{Perceived Capability}), and well-deserved trust (\textit{Justified Trust, Warranted Trust}). Our list of related concepts is not exhaustive, and there could be further concepts that appear outside the domain of our review that we haven't included in our search criteria. 

In summary, we have described the distinction between appropriate trust and related concepts stemming from human beliefs, intentions, and actions. We believe this is one of the first conceptualization in human-AI interaction research to describe, associate, and categorize various concepts in a single framework, which could help reduce the discord among the community on approaching the concept of appropriate trust.

\section{Results of the Systematic Review}
In this section, we review how authors of the included papers define and measure appropriate trust\footnote{We follow the same terminology (appropriate trust/ calibrated trust/ warranted trust etc.) as the authors of the reviewed papers to maintain the consistency.}, what different domains, settings, and tasks they employ, methods for building appropriate trust and the results achieved. 

\subsection{Measures (How to measure Appropriate Trust?)}
Human trust is studied differently based on whether it's conceptualized as a mental attitude \cite{castelfranchi1998principles,falcone2001social}, a belief \cite{jorge2021trust,keren2014trust,zhang2014trust} or a behaviour \cite{castelfranchi2010trust,okamura2020adaptive,yamagishi2015two}. These approaches are typically linked to specific measures which either focus on subjectively measuring attitudes or beliefs, or which look at behavior which demonstrates human trust. As measuring the appropriateness of trust naturally includes measuring trust, we draw the same distinction and divide this subsection into three parts in accordance with Wischnewski et al. \cite{wischnewski2023measuring}: a) Perceived trust, b) Demonstrated trust, and c) Mixed approach. Simply put, we say \emph{perceived trust} is about measuring a person's subjective beliefs, while \emph{demonstrated trust} focuses on their behavior \cite{miller2022we}. While measuring perceived trust is typically done via questionnaires, surveys, interviews, focus groups, and similar reporting tools, demonstrated trust is usually about measuring trust-related behaviors (for instance, in the form of reliance). In demonstrated trust, participants are given the option to use or rely on the system. The underlying assumption is that the more often people use or rely on a system, the more they trust it. 


\subsubsection{Perceived Trust}
Among the papers in our corpus, \textasciitilde40\% measure appropriate trust or related concepts by examining a match between the system's capabilities and the user's trust as a belief. The most common strategy to measure appropriate trust was manipulating a system’s trustworthiness and using self-report scales to compare how self-reported trust adapts to the trustworthiness' levels. For example, Chen et al. \cite{chen2018description} presented participants with either 60\%, 70\%, 80\%, or 90\% reliable systems and measured trust through subjective self-report.  
Similarly, with a within-subjects experimental design, de Visser et al. \cite{de2016almost} had participants interact with a system which trustworthiness' levels were manipulated through its reliability from 100\% to 67\%, 50\%, and, finally, 0\%. Then, the authors used a self-reported trust scale to measure trust which then through comparison with system's trustworthiness provided appropriateness of trust. In the prior examples, manipulating trustworthiness helped the authors to do a before/after comparison. According to Miller \cite{miller2022we} this comparison is crucial for measuring appropriate trust. The authors highlight that without manipulating the trustworthiness of the machine, we cannot establish whether the intervention has correctly calibrated trust. 

We found some authors measured perceived trust by performing a match between the trust ratings and the \textit{static} reliability of the robot or the AI system \cite{ososky2013building,albayram2020investigating,bobko2022human,de2012world,jensen2020role,kraus2019two,lu2019feedback}. However, the match between trust ratings and static reliability may not be perfect. There may be other factors such as appearance or behavior that influence how people rate the trustworthiness of a robot or AI system, even if the system is perfectly reliable. Furthermore, this method does not take into account the dynamic nature of trust. Therefore, we argue that it's difficult to match performance levels with subjective scale ratings.    

\subsubsection{Demonstrated Trust}
We found that only \textasciitilde26\% of studies used behavioral measures for appropriate trust, and we identified three approaches to do so. The most common is \textbf{agreement percentage}, that is the percentage of trials in which the participant’s final prediction agreed with the AI’s correct prediction and cases where participants didn't agree with the AI's wrong prediction \cite{zhang2020confidence,liu2021interactiveexplain,buccinca2021trust,bansal2021does,wang2021explanations,naiseh2021explainrecdesign}. Usually, appropriate trust is seen as a sum of appropriate agreement ratio (human agreement with correct AI predictions) and appropriate disagreement ratio (disagreement with incorrect AI predictions) \cite{liu2021interactiveexplain,coppers2020fortniot,okamura2020adaptive}. Another measure of appropriate trust is related to \textbf{switch percentage} \cite{zhang2020confidence}, that is the percentage of trials in which the participant decided to use the AI’s prediction as their final prediction instead of their own initial prediction. However, it is usually not a standalone measure of appropriate trust and is coupled with other measures. For example, Zhang et al. used a statistically significant interaction between switch and agreement percentages and the AI’s confidence level \cite{zhang2020confidence}. When AI’s confidence level was high and the switch and agreement percentage was high (and vice-versa), then trust was deemed appropriate.

A final method is to measure \textbf{ideal trusting behavior} during the task beforehand and compare to which extent the actual users' behavior matches it \cite{jensen2021trust,herse2021using}. For example, for an experiment where users have to delegate a number of tasks to AI, it is possible to calculate the most optimal number of tasks to delegate to AI in order to achieve the best speed and performance at a given AI's reliability \cite{herse2021using}. The closer users are to this number, the more appropriate their trust in AI is.

\subsubsection{Mixed Approach}
A total of \textasciitilde20\%\footnote{The remaining 14\% of the reviewed papers presented frameworks or theoretical models where no user-study was conducted in which a measure appropriate trust was used.} studies from our corpus used a combination of both self-report measures and behavioral measures to understand appropriate trust. These measures can be categorized into two different subgroups.

The first subgroup includes measures that focus on participants' decisions and compliance with the system's recommendations along with self-reported scales. For example, Wang and Pynadath measured appropriate trust by letting users decide when and when not to trust a low-reliability robot \cite{wang2016explainteam}. They measured self-reported trust by modifying Mayer's scale \cite{mayer1995integrative} and used behavioral measure of compliance as dividing the number of participant decisions that matched the robot’s recommendation by the total number of participant decisions. Accordingly, when both measures matched the reliability of the robot, the trust was considered appropriate. Similarly, Kaniarasu et al. conducted a study where participants rated trust at trust prompts and used buttons to indicate trust changes \cite{kaniarasu2013robot}.  Appropriate trust was measured by examining the degree of alignment of user’s trust with the robot’s current reliability (high or low). Finally, Zhang et al. measured participants’ reliance on AI using two behavioral indicators, agreement frequency and switch-to-agree frequency, as well as via subjective trust ratings \cite{zhang2022complete}. Their diverging reliance and subjective trust ratings results highlight the difference between these two types of measures.

The second subgroup includes measures that examine how participants calibrate their trust over time as they become more familiar with the system's capabilities and policies. For example, Albayram et al. measured how participants calibrated their trust as they grew familiar with the system’s capabilities by using subjective responses and number of images allocated to the automation for pothole inspection by varying automation reliability \cite{albayram2020investigating}. Similarly, de Visser varied the anthropomorphism of the automation to understand trust calibration and appropriate compliance \cite{de2016almost}. By using both subjective ratings and a compliance measure, they measured appropriate trust as the match of a user's trust with the actual reliability of the aid. In both of the previous examples, the researchers manipulated the trustworthiness of the system to measure tan appropriate level of trust. This approach is in line with Miller et al. who states that '\textit{there must be some known or estimated ‘level’ of trustworthiness that is manipulated as part of the evaluation}.' 

\subsubsection{Synopsis}

\noindent 


\noindent In summary, measures of appropriate trust typically involve either a comparison of two different measures: trust of the human and trustworthiness of the system, or they involve some form of agreement metric. The first type naturally involves knowing the trustworthiness of the system. Trustworthiness can be defined as absolute (e.g. the system is correct or not) or relative (the system gets better/worse over time). Although the first might give more insight into how good the system is, it does mean the AI needs to be either wrong or right, which needs to be known. The relative measure allows for an easier comparison, as appropriateness is just about whether trust moves up or down in the same direction as trustworthiness. However, if trust is low for a nearly perfect system and slightly higher but still low for a perfect system, it is still inappropriate despite moving in the correct direction.

Comparing trustworthiness with trust naturally also involves measuring trust. In this also, two methods can be distinguished. The first is subjective and behavioural measures based on questionnaires, and the second is on actions. The main disadvantage of questionnaires is that outcomes can be difficult to directly compare with trustworthiness, while it is easier to establish if reliability is correct. On the other hand, questionnaires better capture the concept of trust as a nuanced belief, as reliance behaviour could be caused by more than just high trust. This is also reflected in the differences between behavioral and subjective scales that can occur when both are used \cite{zhang2022complete}. This highlights the disadvantage of seeing appropriate trust in terms of an agreement metric; this is, by definition, about reliance behaviour and often imposes constraints on the type of human-AI collaboration. Given the limitations of most current measures, the option to use different methods simultaneously has the opportunity to offer a more nuanced result. Which mix is the best might depend highly on the collaboration between the human and AI.

An example of simultaneous use of different methods is (a) the use of validated questionnaires to measure perceived trust combined with (b) behavioural measures to measure demonstrated trust could offer a more insightful measurement than use of one alone \cite{wischnewski2023measuring}. The underlying assumption is that these measures provide an accurate understanding of human's trust. However, as human trust is a multi-dimensional concept its measurement based on scales or behaviour might not provide its complete understanding \cite{spain2008towards}. For example, behavioral measures are context-specific and may not generalize well across different situations and subjective measures may involve participants' individual biases or the willingness to disclose their true feelings. 
Therefore, we propose the next steps in determining how to measure appropriate trust should be to examine combination of measures other than perceived or demonstrated trust. These measures can include personality traits \cite{freitag2016personality}, past experiences \cite{goudge2005can}, social norms \cite{tielman2019deriving}, and cultural values \cite{zhu2018importance}, and how these measures can differ across different contexts and populations. The importance of the context of the task and domains of the study for measurements highlights a need to explicitly define and describe these for studies in appropriate trust.  

\subsection{Tasks}
In this section, we describe the tasks and domains observed in the corpus of this review. We cluster these tasks around distinguishing characteristics which emerged. 

We grouped all studies into different application domains to get an overview of the tasks. In enumerating the domains seen within our corpus of papers (See Table \ref{tab:domains-tasks}), we observe that military operations, transport, and domain agnostic applications are the most common in appropriate trust research. On a more granular level, we see that tasks such as automated driving (n = 14), prediction and classification (n = 14), and reconnaissance (n = 8) are most commonly given to users. Human-AI collaborative tasks such as working in a military environment with humans (e.g., \cite{wang2021explanations}) and teaming for military missions \cite{okamura2020adaptive,tolmeijer2022amoral} are the particularly preferred cases of the reviewed articles. The popularity of military and transport application fields within the study of appropriate trust could follow from the more severe risks associated with the incorrect use of technology in those settings.
 
\begin{table}[h]
\begin{tabular}{|l|l|}
\hline
    Domain            & Tasks   \\ \hline
    Military          & \begin{tabular}[c]{@{}l@{}}Object recognition \cite{lu2019feedback, jensen2020role, choo2022detecting}, Prediction \cite{bobko2022human}, Reconnaissance \cite{wang2016explainteam, guo2021modeling, visser2013adaptdelegate, niedober2014influence, jensen2021trust, okamura2020adaptive, yang2017uxtransparnecy, jensen2018initial}, \\ Remote operation \cite{tolmeijer2022amoral, visser2014cues, johnson2021impact}, Search and Rescue \cite{kaniarasu2013robot}, Non-Experimental \cite{ososky2013building, tomsett2020rapid}\end{tabular} \\ \hline
    Transport         & Automated Driving \cite{ekman2017creating, gremillion2016analysis,  wintersberger2019olfactory, khastgir2018informedsaftey, azevedo2020context, kraus2020more, valentine2021designing, ayoub2021investigation, walker2018changes, m2021calibrating, akash2020toward, helldin2013presenting, mirnig2016framework, kraus2019two}, Non-Experimental \cite{samuel2015predictive}    \\ \hline
    Domain agnostic   & \begin{tabular}[c]{@{}l@{}}Classification  \cite{nesset2021transparency, yang2020visual}, Multi-arm trust game \cite{collins2021miscalnecessary}, Object Recognition  \cite{zhang2022complete},  \\ Non-Experimental \cite{schlicker2021towards, chiou2023trusting, hoffman2017taxonomy, jorge2021trust, shafi2017machine, sheridan2019extending, israelsen2021introducing, devisser2020towards} \end{tabular}    \\ \hline
    Healthcare        & Classification \cite{naiseh2021explainrecdesign, naiseh2021nudging, herse2021using, nesset2021transparency},  Meal Design \cite{buccinca2021trust}, Non-Experimental \cite{mcdermott2019practical, liao2022designing, ferrario2022explaintrust, jacovi2021formalizing}  \\ \hline
    IT                & Prediction \cite{ghai2021explainable}, Classification \cite{bansal2021does}, Question Answering \cite{bansal2021does}, Non-Experimental \cite{jentner2018minions}   \\ \hline
    Justice           & Prediction \cite{wang2021explanations, liu2021interactiveexplain}, Classification \cite{bansal2021does}, Question Answering \cite{bansal2021does} \\ \hline
    Sustainability    & Prediction \cite{wang2021explanations}, Disassembly \cite{alhaji2021physicalasurance} \\ \hline
    Gaming            & Prediction \cite{huang2018establishing}, Classification \cite{de2012world}  \\ \hline
    Robotics          & Non-Experimental \cite{charalambous2016development}, Classification \cite{nesset2021transparency} \\ \hline
    Consumer products & Prediction \cite{coppers2020fortniot}  \\ \hline
    Finance           & Prediction \cite{zhang2020confidence}  \\ \hline
\end{tabular}
\caption{Domains and Associated Tasks Across Our Corpus}
\label{tab:domains-tasks}
\end{table}

When analyzing the breadth of user studies included in this review ($ n = 46$, $45.6\%$ between, $32.6\%$ within subject, $15.2\%$ mixed design), we see a number of patterns emerge in the characteristics of the tasks assigned to participants. We group those characteristics along the dimensions of risk, dynamism, and users' expertise. Interestingly, only three studies \cite{naiseh2021explainrecdesign, valentine2021designing, niedober2014influence} preform a non-controlled experiment, relying on think-aloud sessions, co-design, and interview sessions. They targeted medical, mobility, and military experts for interaction design. To some extent, this does suggest a lack of space within appropriate trust research for the voices of users and stakeholders, and little input on its design processes on their part. 

\subsubsection{Risk}
We highlight risk as an integral part of experimental set-ups, as vulnerability is a key element of trust \cite{lee2004trust, mayer1995integrative}. Yet, it can be overlooked in studies of human-computer trust. We differentiate between explicit and implicit risk using the criteria proposed by Miller \cite{miller2022we}. In these criteria, trust is characterized by the presence of vulnerability and stakes, which introduce a downside to inappropriate trust. The user must be aware of these stakes throughout the experiment, so that actions can be adjusted to accommodate risk. We see that $78.3\%$ of studies include an element of risk in their design. This element is largely implemented in one of two ways; simulated through points gained and lost ($45.7\%$); or incentivized through performance-based pay bonuses ($21.7\%$). Only one study \cite{alhaji2021physicalasurance} used a task which was risky in the experimental setting itself, namely disassembling traction batteries in a recycling context.

The remaining papers rely on the understood risk of a given task (automated driving and remote operation) in the real world to assume users would engage realistically with their experiment \cite{tolmeijer2022amoral, wintersberger2019olfactory, liu2021interactiveexplain, visser2014cues, johnson2021impact}, or do not discuss risk in their methodology \cite{coppers2020fortniot, yang2020visual}. Given the importance of risk to trust, it is difficult to argue that users in such studies demonstrated trust at all, with no consequences attached to over- and under-trust, users may rely on, and positively perceive a system regardless of its actual trustworthiness.  

\subsubsection{Dynamism}
The next element of task design we analyzed is dynamism, that is, changes in Human-AI trust over time informed by the history of interaction \cite{hoff2015trust}. Specifically, we investigate whether studies measure trust levels at multiple points, thus accounting for this dynamic aspect of trust. Across all studies, we find that $63\%$ measure trust more than once throughout the task. In cases of automated driving tasks, this can sometimes even be a continuous measure of trust derived from driver behaviour \cite{akash2020toward, wintersberger2019olfactory, khastgir2018informedsaftey}. Meanwhile, a third of studies measured trust only once throughout the experiment, reducing the complexity of the trust relationship to one snapshot.

Moreover, most of the studies reviewed were either laboratory-based which used simple tasks or theoretical models, which further fails to reflect real-world scenarios. Thus, generalizability of these findings to more complex and dynamic real-world situations is uncertain.

\subsubsection{Participant Expertise}
Overall, $67.4\%$ of studies recruited non-expert participants, because often researchers design tasks so that the participant pool felt equally qualified to complete them without any specific training \cite{bansal2021does, wang2021explanations, azevedo2020context, zhang2020confidence, collins2021miscalnecessary, buccinca2021trust}. Recruitment of non-experts also occurred for the tasks that could require more specialized knowledge, such as military-related tasks \cite{kaniarasu2013robot}. The main reason could be that candidates with required expertise are not available and/or are not easily found. This claim can be supported by the fact that all automated driving studies recruited licensed drivers to their experiments, while only three non-automated-driving user experiment studies recruited expert participants \cite{ghai2021explainable, naiseh2021nudging,tolmeijer2022amoral}. Given that a users' perception of their expertise can affect the extent to which they trust and rely on the automated system \cite{zhang2022complete}, participant expertise should more closely align with the expected expertise of the end user, for more realistic results.

\subsection{Methods for building appropriate trust (How to achieve it?) }
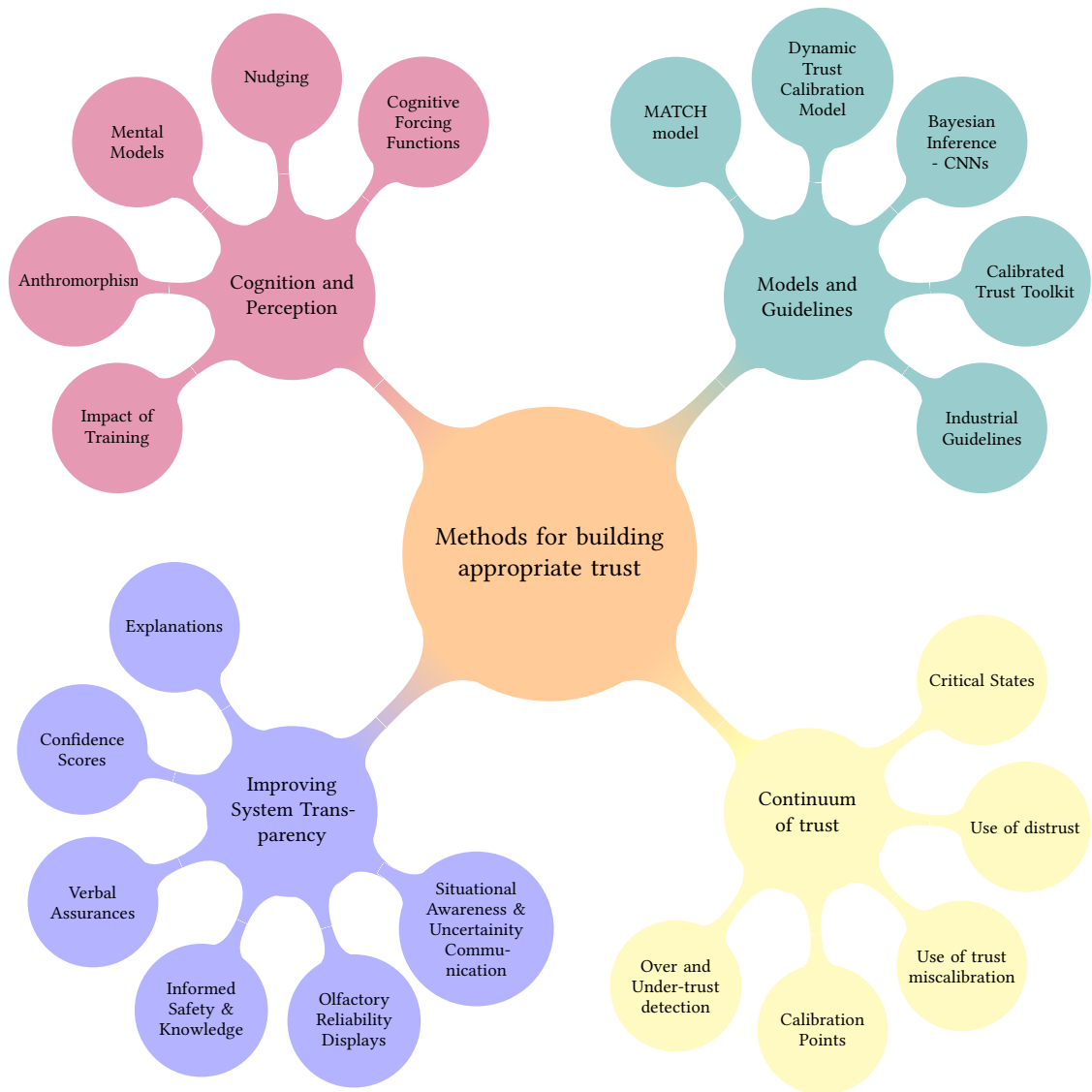
\begin{figure}[h]
    \centering
\begin{tikzpicture}[mindmap, grow cyclic, every node/.style=concept, concept color=orange!40, 
	level 1/.append style={level distance=5cm,sibling angle=90},
	level 2/.append style={level distance=3cm,sibling angle=41},]
\node{Methods for building appropriate trust}
child [concept color=blue!30] { node {Improving System Transparency}
	child { node {Explanations}}
	child { node {Confidence Scores}}
    child { node {Verbal Assurances}}
	child { node {Informed Safety \& Knowledge}}
	child { node {Olfactory Reliability Displays}}
	child { node {Situational Awareness \& Uncertainity Communication}}
}
child [concept color=yellow!30] { node {Continuum of trust}
	child { node {Over and Under-trust detection}}
	child { node {Calibration Points}}
    child { node {Use of trust miscalibration}}
	child { node {Use of distrust}}
    child { node {Critical States}}
}
child [concept color=teal!40] { node {Models and Guidelines}
	child { node {Industrial Guidelines}}
	child { node {Calibrated Trust Toolkit}}
	child { node {Bayesian Inference - CNNs}}
	child { node {Dynamic Trust Calibration Model}}
	child { node {MATCH model}}
}
child [concept color=purple!40] { node {Cognition and Perception}
	child { node {Cognitive Forcing Functions}}
	child { node {Nudging}}
    child { node {Mental Models}}
	child { node {Anthromorphism}}
	child { node {Impact of Training}}
};
\end{tikzpicture}
\caption{Overview of the different appropriate trust building methods adopted in the articles from our corpus.}
    \label{fig:categories}
\end{figure}

In this section, we describe what different approaches were taken towards achieving appropriate trust in the reviewed corpus. A categorization of the methods revealed four broad categories: (1) Improving system transparency, (2) Cognition and perception, (3) Models, guidelines, theories and frameworks, and (4) Relational framing and continuum of trust. These are further shown in Figure \ref{fig:categories}.

\subsubsection{Improving System Transparency}
The first category of methods aims to achieve appropriate trust by adding transparency to systems. 
About 52\% of articles in our corpus target improving transparency of the system to build appropriate trust, \textit{i.e.}, informing users about the specific capabilities and limitations of AI. This indicates that there is a common assumption that improving system transparency can help the human user to better decide when to trust or distrust the AI system. 

One way transparency is improved is through providing \textbf{Explanations}. Explanations focus on the inner-workings of the AI systems (n = 16), appearing either for every AI recommendation \cite{zhang2020confidence,zhang2022complete,tolmeijer2022amoral,ghai2021explainable} or under specific circumstances. For example, Adaptive Explanations by Bansal et al. appear only for the predictions where the AI is quite confident and are absent for the low confidence predictions as a way to avoid human over-trust in the latter case \cite{bansal2021does}. This explanation method was found to be effective in trust calibration, as here the AI system adjusts to the user's attitude and behaviour following the signs of over- and under-trust. To further mitigate over-trust, Lakkaraju et al. call for designing explanations as an Interactive Dialogue where end users can query or explore different explanations for building appropriate trust \cite{lakkaraju2020fool}.
   
Another way to instill transparency is through \textbf{Confidence Scores} of the AI models to align the user’s trustworthiness perception of the system with the actual trustworthiness (n = 12). These scores reflect the chances that the AI is correct, thus relating to its competence and capability. According to Zhang et al., confidence scores are a simple yet effective method for trust calibration \cite{zhang2020confidence}. However, it does not necessarily improve AI-assisted decision-making \cite{bansal2021does}. Furthermore, confidence scores are not always well calibrated in ML classifiers \cite{nguyen2015deep} which can lead to inappropriate trust. 

A combination of explanations and confidence scores has been used for appropriate trust as well under the term of \textbf{Informed Safety \& Knowledge} in relation to autonomous vehicles  \cite{khastgir2018informedsaftey}. The confidence scores informed the drivers of the vehicle's safety. At the same time, the explanations were provided to demonstrate the vehicle's knowledge of any maneuver, enabling the drivers to adjust their level of trust in the system appropriately.

Similar to confidence scores, \textbf{Uncertainty Communication} (n=3), \textit{i.e.}, emphasizing the instances when AI is ``unsure'' of a prediction or does not have a definite answer, can also calibrate trust. For example, an AI agent can yield back the full control to humans and explicitly indicate that it does not ``know'' the solution \cite{tomsett2020rapid}. The results of this method demonstrate that it helps users to spot flows in the reasoning behind the AI predictions and when AI is ``unsure'' about them, and consequently rapidly calibrate their trust.

While confidence scores and uncertainty communication come mostly in a form of a text message, their more anthropomorphized counterpart is \textbf{verbal assurances}. Within this method of transparency, the system verbally indicates to the users what it can and cannot do in a form of promises \cite{albayram2020investigating,alhaji2021physicalasurance} or intent \cite{m2021calibrating}. For example, Albayram et al.'s results show that participants calibrated their trust based on the system’s observed reliability following the promise messages \cite{albayram2020investigating}. Besides written or verbal indicators, odors, presented as \textbf{Olfactory Reliability Displays} \cite{wintersberger2019olfactory}, can also serve to communicate a change in reliability levels of AI for users to calibrate their trust. The authors communicated a change in reliability levels of an automated vehicle simulator using two odors i.e., lemon for a change to low and lavender for a change to high reliability. Their results indicate that olfactory notifications are useful for trust calibration.

Providing more information about not only the AI capability, but also about the task and the context, or in other words, \textbf{Situational Awareness Communication}, can provide transparency to achieve appropriate trust \cite{johnson2021impact,azevedo2020context}. For example, Azevedo et al. showed that with activation of different communication styles to encourage or warn the driver about Situational Awareness (SA) when deemed necessary helps in calibration of trust \cite{azevedo2020context}. Similarly, Johnson et al.'s results show that warning drivers about SA is effective at increasing (decreasing) trust of under-trusting (over-trusting) drivers, and reducing the average trust miscalibration time periods by approximately 40\%.

Studying various methods of improving system transparency for building appropriate trust in AI systems can provide valuable insights. Overall, these works show the value of understanding system transparency and how it can be increased in multiple ways. Some of the most common examples are confidence scores and explanations. However, we also see some unique solutions, such as using olfactory displays or verbal assurances. All these solutions seem promising for improving system transparency, but that communicating system uncertainty or providing real-time situational awareness helps is only sometimes a given. This shows that there is still much to gain, especially in understanding why an AI system is uncertain or what can help to improve its situational awareness for improving system transparency.

\subsubsection{Cognition and Perception}
Another group of methods to achieve appropriate trust is related to human factors, and accounted for 21\% of the reviewed papers. Several of them focus on the \textbf{users' mental model} of AI \cite{ososky2013building}. The more correct the mental model is, the more likely it is that trust will be calibrated appropriately, which links back to our previous method of increasing transparency. One of the ways to achieve this is through training users how to perform the task and how to collaborate with an AI-embedded system \cite{johnson2021impact,naiseh2021explainrecdesign}. The results show that training that emphasized the shortcomings of the system appeared to calibrate expectations and trust \cite{johnson2021impact}. Another way to build a more correct mental model of AI is to let users observe the system’s performance over time \cite{bansal2021does}. By observing the system performance overtime in Bansal et al. study \cite{bansal2021does}, participants developed mental models of the AI’s confidence score to determine when to trust the AI.

Other human factors are related to \textbf{nudging and cognitive forcing functions}. For example, adding friction in the decision-making process of AI to purposefully slow down its recommendation and providing users a nudge gives them an opportunity to better reflect on the final decision \cite{naiseh2021nudging}. Naiseh et al.'s results show that with a nudging based XAI approach such as, (``\textit{You are spending less time than expected in reading the explanation}."), users can calibrate their trust in AI. Similarly, introducing cognitive forcing interventions, \textit{i.e.,} not automatically showing AI recommendations but on-demand or with forced wait can significantly reduce over-reliance compared to the simple explainable AI approaches \cite{buccinca2021trust}. 


Another potential method to calibrate trust through understanding human factors was proposed by Johnson et al. \cite{johnson2021impact}. The authors gave participants trust calibration \textbf{training} about task-work and teamwork before the task. Their results show that training that emphasized the shortcomings of the autonomous agent appeared to calibrate
expectations and trust. Lastly, the characteristics of an AI-embedded system, notably its degree of \textbf{anthropomorphism} contributes to appropriate trust \cite{de2012world}. The results showed that increasing the humanness of the automation increased trust calibration \textit{i.e.}, compliance rates matched with the actual reliability of the aid on increasing humanness.


In synopsis, studying cognition and perception can help us to better understand how people interact with AI systems and how they form impressions of AI systems. Also, studying the mental processes involved in perception, learning, reasoning, and decision-making can help us in designing for appropriate trust in AI systems.

\subsubsection{Models and Guidelines} 
Theoretical foundations can provide insights into how to establish appropriate trust in human-AI interaction (n=12) \cite{schlicker2021towards,israelsen2021introducing,sheridan2019extending,johnson2021impact,visser2014cues,mirnig2016framework}. One example is using models and frameworks to understand how the actual and perceived trustworthiness of AI systems relate to each other. Several papers use different models to explain this relationship and suggest ways to improve it. For instance, Schlicker et al. use two models from organizational psychology to identify factors that influence the match between how trustworthy the system is and how trustworthy the user thinks it is \cite{schlicker2021towards}. Similarly, Israelsen proposes a three-level model that compares the user's and AI's abilities, analyzes the user's past experiences with similar systems, and measures the user's willingness to depend on the system \cite{israelsen2021introducing}. Some similarities between the theoretical models we reviewed are that they often try to explain how the user's perception of the AI system's trustworthiness is influenced by various factors, such as the system's performance, reliability, transparency, feedback, and context. These factors can help us understand how people interact with AI systems. By understanding these factors that influence trustworthiness, we can design AI systems that can be appropriately trusted \cite{mehrotratiis}.

Another type of model focuses on the \textbf{communication of trustworthiness cues} in AI systems. For example, Liao and Sundar \cite{liao2022designing} proposed the MATCH model for responsible trust, which describes how trustworthiness should be communicated in AI systems through trustworthiness cues. With their model, they highlight transparency and interaction as AI systems’ affordances
for designing trustworthiness cues. Apart from communicating trustworthiness cues, some authors studied building appropriate trust by allowing for real-time trust calibration \cite{akash2020toward,shafi2017machine,guo2021modeling}. For example, Shafi \cite{shafi2017machine} provided a parametric model of machine competence that allowed generating different machine competence behaviors based on task difficulty to study \textbf{trust dynamics} for real-time trust calibration. Furthermore, Guo and Yang modeled trust dynamics using \textbf{Bayesian inference} when a human interacts with a robotic agent over time \cite{guo2021modeling}. Here, based on the real-time trust values, a human can calibrate its trust in the robot.

Unlike theoretical models, \textbf{guidelines} offer practical design solutions to achieve calibrated trust in AI. For instance, a \textit{calibrated trust toolkit} \cite{valentine2021designing} aids transparent design of autonomous vehicles, analogous to methods in section 6.3.1. These guidelines address post-design implementation, offering a road-map for human factors in industrial robots and trust calibration for robotic teammates.

Besides academic efforts \cite{visser2014cues,vereschak2021evaluate,chiou2023trusting,naiseh2021explainrecdesign,ososky2013building,tomsett2020rapid}, industrial organizations also offer guidelines for designers and developers of AI-embedded systems \cite{amershi2019guidelines,appleHumanInterface,ibm,googlepair}. These guidelines are often recommendations or best practices that are developed to help people make informed decisions or take specific actions. The majority of \textbf{industrial guidelines} in the field of human-centred AI focus on building users' trust rather achieving appropriate trust (or related terms discussed in Section 5), and only one, Google PAIR guidebook \cite{googlepair}, provides key considerations for users' trust calibration. Examples of their key considerations are telling users what the system can not do, tying explanations to user actions, and considering the risks of a user trusting a false positive or negative. Overall, the key considerations outlined in the Google PAIR guidebook emphasize the importance of effective communication and transparency of the AI models in building appropriate trust in AI systems linking back to the importance of improving system's transparency. 

In synopsis, various theoretical models and guidelines have been proposed to understand the mechanisms around achieving appropriate trust in AI. Theoretical foundations, such as the models and frameworks, provide valuable insights into the factors influencing trustworthiness perception. By examining factors like system performance, reliability, transparency, feedback, and context, we understand how users interact with AI systems, ultimately aiding in designing AI systems that can be appropriately trusted.

Furthermore, models like the MATCH model by Liao and Sundar focus on communicating trustworthiness cues, emphasizing transparency and interaction as essential elements in designing trustworthiness cues in AI systems. Like the one proposed by Shafi, real-time trust calibration models offer insights into how trust dynamics can be managed during human-AI interactions, allowing for adjustments based on task difficulty and performance. In addition to theoretical models, practical guidelines play a vital role in achieving calibrated trust in AI. These guidelines offer actionable recommendations for designers and developers, ensuring that AI systems align with their original design intent. It is worth noting that industrial organizations also contribute to this field, offering guidelines that often focus on building users' trust but increasingly recognize the importance of achieving appropriate trust through effective communication and transparency, as emphasized by the Google PAIR guidebook.

A nuanced approach is crucial in designing trust models for AI systems, considering the intricate interplay of various factors influencing trustworthiness. Likewise, when confronted with many guidelines on trust in AI, tailored selection and adaptation are crucial to ensuring that the chosen guidelines align closely with the unique context, objectives, and stakeholders of the AI system under consideration. Therefore, designing a comprehensive model that addresses all aspects is a complex challenge. Similarly, navigating the many guidelines for building appropriate trust in AI systems can be overwhelming. Therefore, it is essential to consider the specific context, domain, and stakeholders involved. Different guidelines may have varying focuses, such as ethics, explainability, or fairness, so selecting the most relevant ones based on the specific requirements and goals of the AI system can help guide the implementation of appropriate trust measures.
\subsubsection{Continuum of trust}

In order to achieve appropriate trust, one has to be able to recognize when it is not there to fix this. Therefore, studying the entire continuum of trust beyond its appropriate level, \textit{i.e.} over-, under-, mis-, and dis-trust, is helpful in achieving it. For example, it can be possible to achieve calibrated trust through fostering both trust and distrust in AI at the same time \cite{mirnig2016framework}. Sensibly placed distrust makes users not agree with the opinion of others automatically, but rather increases their cognitive flexibility to trust appropriately \cite{oswald2017cooperation}. Yet, only 14\% of the reviewed papers look into this. The literature proposes terms like \textbf{calibration points} \cite{mcdermott2019practical} or \textbf{critical states} \cite{huang2018establishing} to classify the situations when the intervention for calibrating trust is needed. The former term is characterized as a way to classify situations in which the automation excels or situations in which the automation is degraded \cite{mcdermott2019practical}. The later is characterized by the situations in which it is very important to take a certain action such as an autonomous vehicle detects a pedestrian \cite{huang2018establishing}. In both of these situations, a mismatch can occur between levels of performances and expectations, which would allow users to reflect whether their trust levels are appropriate or not. 

Generally, we find that the reviewed papers mostly rely on analyzing human behaviors to determine whether trust needs to be calibrated. For example, states of over- and under-trust are inferred from monitoring the user’s reliance behavior rather than subjective trust measures \cite{okamura2020adaptive}. Collins and Juvina propose to watch out for any behaviors that can be considered as exception out of principle of trust calibration (appropriately calibrated trust) to understand better long-term trust calibration in dynamic environments  \cite{collins2021miscalnecessary}. In their study with a multi-arm trust game, during critical states, users unexpectedly changed their trust strategy, tending to ignore the advice of the previously trusted AI advisors and leaning more towards the previously non-trusted ones. One of the unique findings from this work was that a) trust decays in the absence of evidence of trustworthiness or un-trustworthiness and b) perceived trust necessity and cognitive ability are important antecedents on the trustor's side to detect cues of trustworthiness.

The previous example teaches us that trust calibration is a complex process that requires a nuanced understanding of the context and user behavior, and that the ability to adapt and change trust strategies in response to changing situations is an important aspect of successful trust calibration. Similar to Collins and Juvina, Tang et al. \cite{tang2014distrust} explicitly used distrust behaviors by leveraging data mining and machine learning techniques to model distrust with social media data. Distrust was conceptualized such that it can be a predictor of trust and of the extent to which it is mis-calibrated. Lastly, one paper relied on physiological markers such as gauge behaviour from a eye tracker coupled with the rate of reliance on AI and compared it with the system's capability to identify if trust is mis-calibrated \cite{azevedo2020context}.

In conclusion, there are various approaches adopted by the authors ranging from examining behavior and performance to studying distrust and trust mis-calibration for building appropriate trust. Authors have proposed over- and under-trust detection, calibration points, and critical states to study appropriate trust through the continuum. Furthermore, studies on distrust have shown that it can play a critical role in trust calibration, and trust mis-calibration can be used to understand long-term trust calibration in dynamic environments.

\subsection{Results of calibration interventions}
In this subsection, we provide a general overview of the findings of the reviewed papers. In particular, we focus on the results of applying the methods for building appropriate trust described in Section 6.3. 

From the categories of methods described in this section, improving system transparency was the most common. Most papers supported the hypothesis that transparency facilitates appropriate trust in a system. For example, it was found that uncertainty ratings \cite{tomsett2020rapid}, confidence scores \cite{zhang2020confidence}, providing explanations \cite{bansal2021does,lakkaraju2020fool,nesset2021transparency}, and reliability and situational awareness updates \cite{azevedo2020context} improved appropriate trust in a system. However, other papers add some nuance to this conclusion. For instance, Bansal et al. found that explanations increased the human's acceptance of an AI's recommendation, regardless of its correctness \cite{bansal2021does}. Furthermore, Wang \& Yin found that only some of their tested explanations improved trust calibration, indicating that not all explanations are equal \cite{wang2021explanations}. Lastly, though confidence scores can help calibrate people's trust in an AI model, Zhang et al. (2020) found that this largely depends on whether the human can bring in enough unique knowledge to complement the AI's errors \cite{zhang2020confidence}. These results highlight that further research is necessary to study exactly what methods of increasing transparency are useful to facilitate appropriate trust, given the context of the interaction. We believe opportunities lie in exploring how diverse factors such as user expertise, task complexity, and the type of explanation influence trust calibration. This could involve controlled experiments that manipulate different transparency elements to pinpoint their individual and combined effects on trust. 

Improving system transparency had mixed results for building appropriate trust, and leveraging human cognition and perception for trust calibration yielded the similar results. For example, Riegelsberger et al. found that changes in how a system interacts with the user impacted users' perception of trustworthiness. \cite{riegelsberger2005}. Similarly changing the interaction with the system, Buçinca et al. found that cognitive forcing functions\footnote{Interventions implemented during decision-making to disrupt heuristic reasoning and prompt analytical thinking such as on-demand explanation or forced waiting for output \cite{lambe2016dual}.} reduced over-reliance on AI. However, the performance of human+AI teams was worse than the AI alone with these functions \cite{buccinca2021trust}. Other than the use of cognitive forcing functions to compel people to engage more thoughtfully with AI systems, Naiseh et al. found that nudging can also help users become more receptive and reflective of their decision possibly leading to appropriately trusting the AI system \cite{naiseh2021nudging}. As nudging and cognitive forcing functions target cognitive and perceptual mechanisms for building appropriate trust, the effectiveness of training is also intricately linked to the these mechanisms. For example, two studies showed that teams receiving the calibration training reported that their overall trust in the agent was more robust over time \cite{johnson2021impact,naiseh2021explainrecdesign}. Based on these findings, it is crucial to focus on developing interventions that promote analytical cognitive thinking to foster appropriate trust in AI systems.

The appearance of a system plays a significant role in shaping how humans perceive and mentally process its attributes, which in turn impacts their levels of trust in the system. For example, Jensen et al. discovered that a system with a more human-like appearance was perceived as more benevolent, but this did not lead to differences in trust in behavior leading to unsupported trust calibration \cite{jensen2020role}. Similarly, both Christoforakos et al. \cite{christoforakos2021can} and de Visser et al. \cite{de2016almost} found that more human-like systems were considered more trustworthy, but didn't help in trust calibration. These results highlight that the human-likeness strategies for building appropriate trust have been challenging so far. Although it seems clear there is some effect of appearance on trust, how to use this properly to ensure the appropriateness of trust remains an open question.

So far we have looked at results of the trust calibration interventions related to improving system transparency and understanding human cognition and perception including human-likeness. Distinct from these methods, understanding the continuum of trust was also helpful in certain cases for building appropriate trust. For instance, calibration points and critical states prompted users to adjust their trust in the system by facilitating specific moments of engagements \cite{mcdermott2019practical,huang2018establishing}. Furthermore, detecting over- and under-trust was critical in providing trustworthiness cues to the user in calibrating their trust levels. However, the use of these cues was found to not necessarily improve the performance of the human-AI teams \cite{okamura2020adaptive}. Finally, miscalibrated (i.e., over- or under-) trust \cite{collins2021miscalnecessary} and distrust \cite{kraus2020more} were also promising to calibrate human trust in the system in certain situations such as under conditions of increased trust necessity. Miscalibration affected interactions with new trustors, as a reputation for past trustors preceded the entity, causing potential new trustors to approach with caution \cite{kraus2020more}. Therefore, understanding continuum of trust through user studies can help in building appropriate trust which can improve the human-AI team performance and helpful in trust repair. In particular, opportunities lie in conducting more empirical studies investigating trust development over time with different contexts and how this impacts human decision-making.

In summary, the methods applied in the selected papers yielded mixed results. On the one hand where improving system transparency and understanding human perception and cognition had an impact on appropriateness of trust but on the other hand it did not improve the human+AI joint performance. Similarly, studying the continuum of trust helped in fostering appropriate trust but it also failed to improve human-AI team performance as well as in repairing trust. Overall, it remains complicated to find one-size fits all solution for building appropriate trust in AI systems. Therefore, we recommend that future researchers give careful consideration to a) how they define appropriate trust, b) specify what do they mean by it, c) how they conceptualize their measures and d) avoid using related concepts in particular.

\section{Discussion}
In this systematic review, we have discussed the (a) history of appropriate trust, (b) difference and similarities in concepts related to appropriate trust, (c) a BIA mapping to understand commonalities and differences of related concepts, (d) different methods of developing appropriate trust, as well as (e) results of those methods. In this section, we reflect on our findings by providing critical insights on elaborating key challenges and open questions. Furthermore, we provide some novel perspectives on understanding appropriate trust and finally acknowledge the limitations of this work.
\subsection{Key Challenges}
With appropriate trust constituting a central variable to the appropriate adoption of AI systems, different approaches have been taken to understand it. Our aim with this study was to provide an overview of the field's current state. In doing so, we reflected on our findings and found some challenges that exist in our way of understanding this research area. In this sub-section, we elaborate on the aforementioned key challenges, how to overcome possible limitations and summarize critical points with research opportunities for future work. Our identified key challenges are:
\begin{enumerate}
    \item Discord and diversity in concepts related to appropriate trust such as calibrated trust, justified trust, responsible trust etc. 
    \item A strong focus on appropriate trust in capability, leaving out other aspects of trust such as benevolence and integrity \cite{jorge2021trust}.
    \item The issues involved in adequately measuring appropriate trust.
\end{enumerate}

\subsubsection{Discord and diversity in understanding appropriate trust}
From the analysis of the reviewed definitions of appropriate trust, we identify 3 major challenges for the current theoretical discourse on the topic.
Firstly, as seen in Section 5, there is no uniform understanding on what appropriate trust is: some papers define appropriate trust based on system performance or reliability \cite{yang2020visual,okamura2020adaptive,ososky2013building,niedober2014influence,walker2018changes}, some relate it to trustworthiness and beliefs \cite{jorge2021trust,danks} and some base it on calculations\cite{jensen2021trust,coppers2020fortniot,wang2021explanations}. Such a variety of the appropriate trust definitions stems from different understanding of what ``the right amount of trust" implies. The common denominators of having various definitions of appropriate trust can be linked to: (a) the context in which it is studied often differs from one study to another, (b) the multidimensional nature of trust, often associated with attitude or subjective beliefs, adds complexity to understanding appropriate trust, and (c) different academic fields approach the study of trust in unique ways, leading to divergent interpretations of appropriate trust. For example, in HRI domain trust is often linked to robot's performance \cite{kaniarasu2013robot} whereas in Psychology it is commonly linked to understanding social and interpersonal aspects \cite{rempel1985trust}.

In addition to the variety of definitions of appropriate trust, we also found that the literature proposes various related concepts \footnote{From our understanding, a "concept" is a general idea representing a category, while a "definition" is a precise statement that clarifies the meaning of a term or concept.} (See Table \ref{tab:definitions}), sometimes used interchangeably in the discourse about appropriate trust \cite{wintersberger2019olfactory,bansal2021does,okamura2020adaptive}.
For example, we would like to especially stress the difference between appropriate trust and another most used related construct - calibrated trust.  
Although the logical formulation of the two concepts is similar as shown in the BIA mapping in Figure \ref{fig:trust_concept_map}, trust calibration requires a process. In contrast, appropriate trust is the maintained state of the calibrated trust over a series of interactions. This conceptual overlap raises questions about the precise boundaries and distinctions between these concepts and highlights the need for a more refined and standardized conceptual framework. 

These challenges surrounding understanding appropriate trust emphasize the significance of shaping our research agenda in this domain. To address the need for consensus among researchers, in this work we proposed a framework that explicitly defines appropriate trust and its boundaries. Our framework consider multiple dimensions, such as system capability, trustworthiness, beliefs, and task requirements while accounting for contextual variations. Moreover, we made an attempt to clarify the relationships between appropriate trust and related concepts, establishing clear definitions and boundaries to facilitate meaningful discussions and avoid conceptual confusion. By addressing these challenges and shaping a coherent research agenda, we can advance our understanding of appropriate trust and its implications for various domains.

\subsubsection{Prominent focus on system's capabilities in definitions}
The majority of appropriate trust definitions or its related concepts focus on the capability or ability of an agent. Here, appropriate trust is the alignment between the perceived and actual capabilities of the agent by the human \cite{yang2020visual,okamura2020adaptive,liu2021interactiveexplain}. Much of previous research has looked at `ability' as the core factor of establishing trust \cite{jorge2021trust,mehrotra2021modelling}, which bring the focus upon the engineering aspect of trustworthiness. However, we view trustworthiness as more than just ability. Our interpretation of trustworthiness can be enhanced when we not only focus upon agent capabilities but also on understanding other factors such as integrity and benevolence \cite{mayer1995integrative,parasuraman2004trust} or process and purpose \cite{lee2004trust}. 

Hoffman et al. state that ``a thorough understanding of both the psychological and engineering aspects of trust is necessary to develop an appropriate trust model" \cite{hoffman2017taxonomy}. Our examination of the psychological aspects of trust in human-AI interaction has revealed a need for improvement in the existing literature regarding modeling the integrity and benevolence of an AI agent toward a human as highlighted by Ulfert et al. \cite{ulfert2023shaping}, Mehrotra et al. \cite{mehrotratiis} and Jorge et al. \cite{jorge2021trust}. 
Mayer et al. \cite{mayer1995integrative} propose that the effect of integrity on human trust will be most salient early in the relationship, before the development of meaningful benevolence \textit{i.e.,} 
X has disposition to do good for Y \cite{deutsch1958trust}. Therefore, we pose that it is important to first investigate how humans perceive AI system's integrity and how to model this relationship for fostering appropriate trust in AI system. 
Then it becomes vital to study the effect of perceived benevolence on trust as it increases over time as the relationship between the parties develops \cite{mehrotra2021modelling}. Throughout, the perceived ability of the system remains important. However, we pose it is crucial to not forget these other factors in research on appropriate trust.

\subsubsection{Adequately Measuring Appropriate Trust}
While analyzing our corpus, we encountered common issues with appropriate trust measurements identified by Miller \cite{miller2022we}. These issues include the absence of risk and vulnerability elements in user studies; overlooking instances of under-trust; uncertainty regarding the extent to which behavioral experiments can capture trust; the robustness of single/multiple-item questionnaires in capturing changes in trust levels over time; reliance on agreement/disagreement with model predictions without considering discrepancies in human goals; and the use of appropriate situational awareness as a proxy for trust.

First, we found some papers in our corpus \cite{kaniarasu2013robot,coppers2020fortniot,schlicker2021towards,alhaji2021physicalasurance,naiseh2021explainrecdesign} which had little or no element of risk in the task design. We posit that in a questionnaire, survey, or field study it is crucial that participants have experienced or currently experience vulnerability to the possibility of the AI system failing. Trust cannot exist without the element of risk, and participants must have a personal stake in the situation. Including risk and vulnerability factors allows researchers to evaluate the trustworthiness of systems or services accurately.

Second, we observed some articles focused on capturing over-trust in AI \cite{wintersberger2019olfactory,wang2016explainteam,buccinca2021trust,jentner2018minions,israelsen2021introducing}, however under-trust was often overlooked. We posit that calibrated trust requires equal consideration of both scenarios. Appropriate trust necessitates equal consideration of both over-trust and under-trust scenarios because a skewed focus on one aspect can lead to sub-optimal outcomes.

Third, it is not clear to what extent behavioral experiments which account for 70\% of experiment designs, especially physiological \& empirical measures, can be used as a proxy to capture trust. While behavioral experiments can offer valuable insights into trust-related behaviors, their ability to fully capture the complexity of trust can be unclear due to simplified environments, artificial motivations, lack of context, limited generalizability, and the subjective nature of trust \cite{erle2020illusory}. 

Fourth, it is difficult to establish whether single/multiple-item questionnaires are robust enough to capture changes in trust levels over time \cite{johnson2021impact,kraus2020more,collins2021miscalnecessary,alhaji2021physicalasurance,khastgir2018informedsaftey}. Also, in almost 40\% of studies trust is measured before and after the user study, though it is not always appropriate to reflect on users’ attitude at such a high level of granularity. A focus on trust dynamics over time as indicated by some studies \cite{kraus2020more,alhaji2021physicalasurance,ayoub2021investigation,guo2021modeling} could be a better approach.  

Fifth, measures of trust related to whether humans agree or disagree with a model prediction are employed in some studies \cite{zhang2022complete,zhang2020confidence,liu2021interactiveexplain,buccinca2021trust}, however what happens when the model targets differ from human goals? Sixth, reliance was often used as a proxy for trust, or even treated as the same thing. As Tolmeijer et al. \cite{tolmeijer2022amoral} highlighted trust in an agent as the belief that “an agent will help achieve an individual’s goal in a situation characterized by uncertainty and vulnerability” \cite{lee2004trust}, while reliance on AI is defined as “a discrete process of engaging or disengaging” \cite{lee2004trust} with the AI system. Finally, some authors \cite{azevedo2020context,johnson2021impact} acknowledge the ambiguity of using appropriate situational awareness as a proxy for measuring appropriate trust in their approach. 
 
In this sub-section, we explored the discord and diversity in concepts related to appropriate trust, including calibrated trust, justified trust, and responsible trust. We also highlighted a prevalent focus on trust in capability, neglecting other important aspects like benevolence and integrity. We found a lack of clear understanding of appropriate trust and identified issues in assessing it. Finally, we have yet to completely characterize how to measure appropriate trust adequately. For example, there is more work to do to fully understand the element of risk or vulnerability, to have a clear distinction between reliance and trust, and an uneven focus on both over- and under-trust. 

\subsection{Open Questions}
While analyzing the text from our corpus, we discovered some open questions on determining whether appropriate trust in AI systems is achieved. First, what to take into account when deciding whether human's trust in the AI system is over-trust or under-trust? From the reviewed articles, this distinction seems to be primarily based on the AI accuracy, \textit{i.e.,} correct or incorrect AI recommendations\cite{yang2020visual,zhang2022complete,zhang2020confidence}. 
We argue this process of determining where the threshold lies in deciding over- or under-trust ca not be solely about making a right or wrong decision; instead it should consider multiple aspects. For example, while accuracy indicates human reliance on the AI system's outputs, it does not capture the nuanced nature of trust. Trust involves more than mere reliance; it encompasses perceived reliability, multiple interactions, transparency, and the belief that the AI system has the user's best interests. For instance, a user may rely on an AI-based navigation system when using it for the first time to reach their destination, leading to 100\% reliance. However, trusting the system 100\% may require interacting with it multiple times in different contexts. Hence, we argue that a comprehensive evaluation of trust should consider a multidimensional approach that incorporates both accuracy and factors related to transparency, interpretability, adaptability, longitudinal interactions, user feedback, and the cognitive and emotional aspects of trust. This broader perspective will enable researchers to understand better when human trust in an AI system gears towards over-trust or under-trust \cite{nooteboom2013trust}.

Second, how to calculate appropriate trust for a task with non-binary decision-making? \textit{i.e.,} when the decisions are non-binary (e.g., price estimation) it is relatively difficult to identify over- and under-trust at regular time intervals. This could be because it involves a continuous scale of possibilities, making it challenging to define clear boundaries for what constitutes over-trust or under-trust. However, when the decisions are binary it is easier to assess trust since one can directly compare the outcomes to the binary decisions (e.g., correct or incorrect). In our analysis, we could not find any articles from the reviewed corpus that clarify how to calculate appropriate trust if the decisions are non-binary. We believe in such cases, it is essential to consider a more nuanced approach that takes into account the specific characteristics of the task and the decision-making process such as by assigning probabilities to different outcomes or decision options.

Third, and relating to the previous point, as AI systems can change overtime, so how can we measure appropriate trust, or even reliance, as they becoming moving targets? Consider the automated vehicle which is highly reliable in dry, clean, weather but whose performance degrades in rainy conditions, forcing the driver to dynamically adjust their trust. We only find mention of this limitation in five of the articles we reviewed. Further, we could not find reviewed articles addressing how periodicity in the trust gain and loss is affected by the task \textit{i.e.,} frequency or regularity with which trust is gained or lost in a task, thus we have limited understanding of trust dynamics in real-world long-term interactions. We postulate that a common reason why we couldn't find articles relating to periodicity of trust is because dynamics of trust development and erosion is itself a complex topic which can impact task performance and efficiency. Hence, we need further research on  generating empirical evidence, insights, and theoretical frameworks to address the gap in knowledge regarding the influence of task frequency and regularity on the periodicity of trust gain and loss. 



\subsection{Novel perspectives}
We found some distinct perspectives on understanding appropriate trust in AI while analyzing our corpus. First, Chiou and Lee argue that the current approach to studying trust calibration neglects relational aspects of increasingly capable automation and system-level outcomes, such as cooperation and resilience \cite{chiou2023trusting}. They adopt a relational framing of trust to address these limitations based on the decision situation, semiotics, interaction sequence, and strategy. They stress that the goal is not to maximize or even calibrate trust, but to support a process of trusting through automation responsivity. We resonate with the perspective put forward by the authors; however, to achieve a higher degree of automation responsivity, human values, societal norms, and conflicts are to be studied and implied in the AI systems.

Second, Toreini et al. suggest that we need to study the locus of trust to understand appropriate trust in the systems \cite{toreini2020relationship}. They raise the questions such as whether we trust the people who developed the system or the system itself. What purpose are the broader organizations serving? Furthermore, the authors acknowledge the limitations of individuals’ capabilities concerning assessing ability and benevolence and propose that individuals accomplish this indirectly by assessing the ability and benevolence of the entity developing the AI. Finally, among the enormous amount of methods and approaches presented in the review, the work by Collins and Juvina highlights the importance of trust mis-calibrations to study appropriate trust \cite{collins2021miscalnecessary}. According to the authors, when the need for trust becomes stronger, individuals may stop trusting their previous trusted partners and instead try to establish trust with those they previously distrusted. Studying these exceptions to the principle of trust calibration might be critical for understanding long-term trust calibration in dynamic environments. We believe that this change in trust tactics which is known in human-human interaction is missing in the human-AI interaction studies. Furthermore, we couldn't find any studies in which humans interact with several AI systems in real life, so this aspect of trust strategies needs to be studied if we wish to learn about how trust mis-calibration can be a useful tool to understand appropriate trust in AI systems.

\subsection{Limitations}
Despite the systematic review's comprehensive analysis of the state of the art in fostering appropriate trust, there are several limitations to this study that need to be acknowledged. 

First, while we included studies from limited disciplines (refer our search string in section \ref{search_string}), it is possible that some relevant studies were missed. Additionally, we only focused on studies published in English, which may have led to language bias. Future reviews should consider including studies in other languages to increase the generalizability of the findings. 


Second, our mapping to concepts related to appropriate trust based on beliefs, desires, and intention is only one of many possible ways to organize such concepts under an umbrella. As such, future research can focus on the development of a clear and concise mapping of these definitions from a multidisciplinary perspective. 

Third, our search period only included papers from 2012 till June 2022 and the research on appropriate trust is growing at a faster pace. Therefore, papers which were published from June 2022 are missing from this review. Finally, the current review only focused on the current state of the art in fostering appropriate trust in AI systems. While the review identified potential research gaps and opportunities, additional research is necessary to develop new approaches and design techniques to better understand the topic.

\subsection{Summary}
This sub-section aims to summarize the current trends, challenges, and recommendations concerning the definitions, conceptualizations, measures, implications of measures, and results for establishing appropriate trust in AI systems. By addressing the evolving trends, inherent challenges, and potential solutions, we aim to enrich the overall understanding of the topic, enabling readers to grasp the broader context and implications associated with building appropriate trust in AI systems.

Our aim with this summary is to provide a well-structured gateway for both experts and newcomers to understand the trends and challenges with an actionable set of recommendations. With these recommendations we make an attempt to connect all the sections of this paper to provide broader context and implications of building appropriate trust in AI.
\begin{longtable}{p{0.1\textwidth}|p{0.3\textwidth}p{0.3\textwidth} p{0.3\textwidth}}
\toprule
\textbf{Section} & \textbf{Current Trends} & \textbf{Challenges} & \textbf{Recommendations}  \\   \midrule 
\textbf{Definitions} & \textbf{(1)} 75.3\% (n = 312) of articles from our corpus which were sought for retrieval did not provide a definition of appropriate trust or a related concept \footnote{Italics is for supplementing the information.}. & \textbf{(1) }A lack of clear definition creates a confusion among readers from different backgrounds.   & \textbf{(1)} Provide a clear definition of appropriate trust or a related concept. \\ \\ & \textbf{(2)} Of the articles, which provided a definition in our final corpus, 25\% (n = 16) of them provided new definitions which were often not related prior works, see Table \ref{tab:definitions}.  & \textbf{(2)} A variety of definitions inherent to multidisciplinary fields without relating it to other fields can cause misunderstanding to the reader. & \textbf{(2)} We need to converge in the future to establish common ground to define \textit{what appropriate trust means in human-AI interaction?}\\
\midrule
 
\textbf{Conceptual-ization} & \textbf{(1)} Many types of appropriate trust concepts are only sometimes explicitly distinguished. 

\textit{For example, the differences between optimal trust, well-placed trust, meaningful trust, justified trust etc., are often unclear and used interchangeably.} & \textbf{(1)} A plethora of concepts related to appropriate trust is causing the HCI community to diverge in multiple ways. 

\textit{This unclear connotation of similar concepts often creates confusion among researchers, especially new graduate students.} & \textbf{(1)} Related concepts which are distinct from the goal of appropriate trust should be defined, distinguished, treated and measured as independent concepts. 

\textit{For example, warranted trust and contractual trust have different goals than appropriate trust.}  \\ \\ & \textbf{(2)} Interchangeable use of Appropriate Trust with Appropriate Reliance  & \textbf{(2)} A core distinction in philosophy, which is often neglected in the empirical HCI literature, regards trust and reliance as distinct concepts. & \textbf{(2)} We propose Hoff \& Bashir distinction \cite{hoff2015trust}, where trust is the belief that “\textit{an agent will help achieve an individual’s goal in a situation characterized by uncertainty and vulnerability}” and Lee and See's reliance distinction \cite{lee2004trust} “\textit{a discrete process of engaging or disengaging}”. \\ \\ &  \textbf{(3)} 38\% of articles in our final corpus conceptualize appropriate trust or related concepts as the measure of alignment between the perceived and actual ability of the system.  & \textbf{(3)} To explore the extent and magnitude of how the trustworthiness properties of machines, beyond their ability, impact trust. For example, what do we mean by integrity of a machine, and how can we measure it?  & \textbf{(3)} We must focus on measuring less studied dimensions of trustworthiness, i.e., integrity and benevolence, to understand human trust in AI systems. \\  &
\\
\midrule
\textbf{Measures} & \textbf{(1)} 40\% (n=26) of articles in our final corpus study appropriate trust in binary decision making tasks \textit{i.e.} to [not] follow an [in]correct AI recommendation. & \textbf{(1)} To develop strategies for building appropriate trust in AI systems that continuously make decisions, such as in price estimation. Also, the potential issues that arise when the AI model targets diverge from human goals. & \textbf{(1)} We need to investigate new measures to assess dynamic trust in practice. For example, we can use situational reference points to keep aligning the goal \cite{chen2008goal}.\\
\midrule
\textbf{Results} & \textbf{(1)} Around 37\% of reviewed articles report the effect of improving system transparency for establishing appropriate trust in human-AI interaction. & \textbf{(1)} A disadvantage of single focus on improving system transparency requires ground truth, which is often not available or there is no really ‘ground’ at all. & \textbf{(1)} Include post-experiment surveys or interviews where the participants can give their impressions on the trustworthiness of the AI Agents.  \\ \\ & \textbf{(2)} In 43\% of the included articles, the objective of the designed task had direct influence on the results of appropriate trust in human-AI interaction. & \textbf{(2)} If the objective of the task to foster appropriate trust in the AI agent is built around improving the fairness of the AI agent then the results will be different compared to objective of improving the accuracy. & \textbf{(2)} Ensure to control initial participant's expectations about the AI system and report results with scientific rigor about how the design of the task may have influenced human trust.\\
\midrule
\textbf{Implication of the measures} & \textbf{(1)} 45\% of articles involving a user-study focused on detecting over trust in AI, under trust in AI systems is often overlooked. & \textbf{(1)} Under-trust in AI systems is a common challenge.  & \textbf{(1)} Investigate and adopt methodologies from social sciences and psychology to study under trust in AI \cite{josang2004analysing}. \\ & \textbf{(2)} Around 10\% of articles in our corpus follow some already established guidelines to design for fostering appropriate trust. & \textbf{(2)} There are multiple guidelines from academia and industrial organizations outlining trust calibration principles that AI-based systems should adopt. However, there is less effort that has been put in translating those principles into practice. & \textbf{(2)} Adopt established guidelines while designing an user study and report if those guidelines did not scaled for the user study. \\ & \textbf{(3)} Locus of trust in the AI systems: are we trusting the people who developed the system is unexplored. & \textbf{(3)} Identify and explore the fundamental correlations between appropriate trust in AI systems and the manufacturers of AI. & \textbf{(3)} Adopt Toreini et al. \cite{toreini2020relationship} recommendation on analyzing factors such as the transparency of the AI development process, the track record of the manufacturer in delivering trustworthy AI, and the level of accountability and responsibility taken by the manufacturer for the AI's outcomes. \\ 

\\
\bottomrule                       
\caption{A detailed summary of current trends, challenges and recommendations based on the results of the systematic review.}
\label{tab:my-table}
\end{longtable}

\section{Conclusion}
Appropriate trust in AI systems is crucial for effective collaboration between humans and AI systems. Various approaches have been taken to build and assess appropriate trust in AI systems in the past. This paper provides a comprehensive understanding of the field with a systematic review outlining different definitions of appropriate trust, methods to achieve it, results of those methods, and a detailed discussion on challenges and future considerations. Through this review of current practices in building appropriate trust, we have identified the challenge for a single definition of appropriate trust and the ambiguity surrounding related concepts such as warranted trust, appropriate reliance, or justified trust. 

Our review has proposed a Belief, Intentions, and Actions (BIA) mapping to study commonalities and differences among different concepts related to appropriate trust. We found three common measurement techniques to measure appropriate trust as Perceived, Demonstrated and Mixed. In addition, multiple domains and associated tasks have been used to study appropriate trust. Furthermore, our analysis of articles revealed four common methods for building appropriate trust such as transparency, perception, guidelines and studying the continuum of trust. 

In synopsis, the review highlights what approaches exist to build appropriate trust and how successful they seem to be. We have discussed the challenges and potential gaps in studying appropriate trust, which presents opportunities for future research such as discord \& diversity in defining appropriate trust or a strong focus on capability. 
Overall, this paper provides (a) a comprehensive overview of the current state of research on appropriate trust in AI by studying measures, tasks, methods, and results of those methods, (b) a BIA mapping of appropriate trust and its related concepts, and (c) a set of recommendations for fostering appropriate trust in AI based on current trends and challenges. With these contributions, we can advance our understanding of designing for appropriate trust in Human-AI interaction taking a step closer towards Responsible AI \cite{IEEE}.
\begin{acks}
This research was (partly) funded by the Hybrid Intelligence Center, a 10-year programme funded the Dutch Ministry of Education, Culture and Science through the Netherlands Organisation for Scientiic Research, grant number 024.004.022 and by EU H2020 ICT48 project "Humane AI Net" under contract 952026. Furthermore, we thank Ewart de Visser, Enrico Liscio and Mohammed Al Owayyed for their contribution in iterations of this paper.
\end{acks}

\bibliographystyle{acm}
\bibliography{main}
\end{document}